\documentclass[aps,prc,amsfonts,preprintnumbers,superscriptaddress,showpacs,nofootinbib]{revtex4}

\usepackage{graphics}
\usepackage{epsfig}
\usepackage{float}

\newcommand{\fet}[1]{\mbox{\boldmath $#1$}}

\newcommand{\beq}{\begin{equation}}
\newcommand{\eeq}{\end{equation}}
\newcommand{\beqa}{\begin{eqnarray}}
\newcommand{\eeqa}{\end{eqnarray}}
\newcommand{\nn}{\nonumber \\ }

\begin{document}

\preprint{FZJ-IKP-TH-2007-26}
\preprint{HISKP-TH-07/26}


\title{Four-nucleon force using the method of unitary transformation}


\author{E. Epelbaum}
\email[]{Email: e.epelbaum@fz-juelich.de}
\affiliation{Forschungszentrum J\"ulich, Institut f\"ur Kernphysik 
(Theorie), D-52425 J\"ulich, Germany}
\affiliation{Universit\"at Bonn, Helmholtz-Institut f{\"u}r
  Strahlen- und Kernphysik (Theorie), D-53115 Bonn, Germany}


\date{\today}

\begin{abstract}
We discuss in detail the derivation of the leading four-nucleon force in
chiral effective field theory using the method of unitary transformation. 
The resulting four-nucleon force is given in both momentum and
configuration space.  It does not contain any unknown parameters and can be
used in few- and many-nucleon studies. 
\end{abstract}


\maketitle


\vspace{-0.2cm}

\section{Introduction}
\def\theequation{\arabic{section}.\arabic{equation}}
\label{sec:intro}

Chiral effective field theory (EFT) is a powerful tool for analyzing the properties of 
hadronic systems at low energies in a systematic and model-independent way. 
It exploits the approximate and spontaneously broken chiral symmetry of QCD 
which governs low-energy hadron structure and dynamics. Over the past years, considerable progress 
has been achieved in understanding the structure of the nuclear force in 
this framework, see
\cite{Bedaque:2002mn,Epelbaum:2005pn} for 
recent review articles. In particular, two-nucleon force (2NF) 
has been worked out and applied in few-nucleon studies upto
next-to-next-to-next-to-leading (N$^3$LO) in
the chiral expansion \cite{Entem:2003ft,Epelbaum:2004fk}. 
Three-nucleon force is currently available at
next-to-next-to-leading order (N$^2$LO) \cite{vanKolck:1994yi,Epelbaum:2002vt},
see \cite{Witala:2006nn,Kistryn:2005aa,Navratil:2007we} for some examples of recent
few-nucleon studies at N$^2$LO. Although the results for most of the investigated
few-nucleon observables look promising, it is
necessary to go to N$^3$LO in order to test the convergence of the chiral
expansion. Increasing the chiral order is also expected to bring new insights
into certain existing puzzles in the 3N continuum such as e.g.~the large
deviations from the data for the differential cross section in some
deuteron breakup configurations, see \cite{Ley:2006hu} for a recent work on this
subject. The full N$^3$LO analysis requires the incorporation of the
subleading contributions to
the 3NF which are currently being worked out. In addition, one has to take
into account four-nucleon force (4NF) which  starts to contribute at this
order. In Ref.~\cite{Epelbaum:2006eu} we already presented the expressions for
the leading 4NF. It is governed by the exchange of pions and the lowest-order 
nucleon-nucleon contact interaction and includes effects due to the nonlinear  
pion-nucleon couplings and the pion self-interactions constrained by the chiral 
symmetry of QCD. The obtained 4NF is local and does not
contain any unknown parameters. 
In this work, we describe in detail the derivation of
the 4NF using the method of unitary transformation
\cite{Epelbaum:1998ka,Epelbaum:2000kv}, for which a new formulation
is presented. This new formulation is considerably simpler than the one given in 
Refs.~\cite{Epelbaum:1998ka,Epelbaum:2000kv} and much more convenient for practical
applications. We furthermore study the effects of the additional unitary
transformations which affect the form of the effective Hamilton operator at
N$^3$LO. The appearance of such transformations is a new feature at this order
in the chiral expansion. 

Our manuscript is organized as follows. In section \ref{sec:MUT} we discuss the
chiral power counting in the few-nucleon sector following the original
formulation by Weinberg \cite{Weinberg:1990rz,Weinberg:1991um}. 
We express the power of the low-momentum
scale in a different form as compared to
Refs.~\cite{Weinberg:1990rz,Weinberg:1991um} 
which is  particularly useful for
applications based on algebraic rather than diagrammatic methods. Combining
this new formulation with the method of unitary transformation we obtain the formal
expression for the effective nuclear Hamilton operator in a compact form. 
In section \ref{sec:N3LO} we use this new formulation to derive the N$^3$LO 
contributions to the effective nuclear Hamiltonian and evaluate the
corresponding four-nucleon operators which give rise to the 4NF. 
We employ a large class of additional unitary transformations acting on the
nucleon subspace of the full Fock space and investigate their impact on 3NFs and 4NFs. 
We end with the summary and outlook. Finally, appendix \ref{app1} contains some
details related to the N$^3$LO contributions involving subleading
vertices while the expressions for the 4NF in configuration space are given in
appendix \ref{app:coord}.

\section{Nuclear forces in chiral EFT using the method of unitary 
transformation}
\def\theequation{\arabic{section}.\arabic{equation}}
\label{sec:MUT}

\subsection{Chiral power counting}
\label{sec:PC}

Power counting is a crucial ingredient of any effective field theory
which  allows to organize various contributions to the scattering
amplitude according to their relevance. Throughout this work, we adopt
Weinberg's power counting which is based on naive dimensional 
analysis\footnote{Notice that alternative counting schemes for contact 
two-nucleon interactions are currently being explored.}.  
Traditionally, chiral power counting
is formulated in terms of topological quantities such as the number of loops
and, in the few-nucleon
sector, the number of participating nucleons and disconnected pieces  
\cite{Weinberg:1978kz,Weinberg:1990rz,Weinberg:1991um}. 
This allows to immediately identify all relevant diagrams at a given order and
appears to be  
particularly useful for calculations based on the Feynman
graph technique. For algebraic approaches such as the method of unitary
transformation \cite{Epelbaum:1998ka,Epelbaum:2000kv} it is more
convenient to express the power counting in a different form which will
be given below. 
For the sake of completeness, let us first recall the derivation of Weinberg's power counting
result following his original work \cite{Weinberg:1990rz,Weinberg:1991um}. 
Consider an arbitrary $N$-nucleon irreducible time-ordered diagram (i.e. the one that does
not contain iterative contributions) without external
pions. For the sake of generality, we allow for cases in which $\tilde N$ out of $N$ 
nucleons do not interact. To obtain the corresponding power $\nu$ of the soft
scale $Q$ one has to count the number of the independent three-dimensional momentum-space
integrations $I - \sum_i V_i$ with $I$ ($V_i$) being the number of internal
lines (vertices of type $i$), the number $d_i$ of derivatives or pion mass insertions at
each vertex, the number $D$ of energy denominators and the number of
phase-space factors associated with exchanged pions:
\beq
\label{start}
\nu = 3 I - 3 \left( \sum_i V_i - 1 \right) - D + \sum_i  V_i \left( d_i - \frac{p_i}{2} \right) - 3 \tilde N\,.
\eeq
Here, $p_i$ denotes the number of pion fields at a vertex of type $i$. 
Notice that we follow the standard convention to define the potential. In
particular, we 
do not count the overall $\delta$-function that enters the
definition of the S-matrix element but do count the additional
$\delta$-functions associated with the noninteracting nucleons. 
To bring Eq.~(\ref{start}) into the standard form one can use the following
topological identities. 
First, the number of intermediate states $D$ can be expressed as:
\beq
\label{eq0}
D = \sum_i V_i -1\,.
\eeq
Secondly, the number of loops $L$ is given by 
\beq
\label{eq1}
L = I - \sum_i V_i + C - \tilde N\,,
\eeq
where $C$ refers to the number of separately connected pieces. Notice that  
each noninteracting nucleon is regarded as a separately connected piece. 
The last identity we need reads
\beq
\label{eq2}
2 I + 2 N = \sum_i V_i (p_i + n_i) + 2 \tilde N \,. 
\eeq
Here, $n_i$ is the number of nucleon field operators at a vertex of type $i$.
Using Eqs.~(\ref{eq0}), (\ref{eq1}) and  (\ref{eq2}), one can rewrite
Eq.~(\ref{start}) in the form that appears in Weinberg's original papers
\cite{Weinberg:1990rz,Weinberg:1991um,Weinberg:1992yk}:
\beq
\label{pow_orig}
\nu = 4 - N + 2 (L - C) + \sum_i V_i \Delta_i \,, \quad \quad
\Delta_i = d_i + \frac{1}{2} n_i - 2\,.
\eeq
There is one subtlety in the above expression which needs to be addressed: 
according to Eq.~(\ref{pow_orig}), the chiral dimension $\nu$ for a given
process depends on the total number of nucleons in the system. For example,
one-pion exchange in the two-nucleon system corresponds to $N=2$, $L=0$,
$C=1$ and $\sum_i V_i \Delta_i =0$ and, therefore, contributes at order $\nu
=0$. On the other hand, the same process in the presence of the third
(spectator) nucleon leads, according to Eq.~(\ref{pow_orig}), to $\nu = -3$
since  $N=3$ and $C=2$.  The origin of this discrepancy lies obviously 
in the different normalization of the 2N and 3N states:
\beqa 
&2N:& \quad \langle \vec p_1 \, \vec p_2 | \vec p_1 {}' \,  \vec p_2 {}' 
\rangle = \delta^3 (\vec p_1 {} ' - \vec p_1 \, ) \, 
\delta^3 (\vec p_2 {}' - \vec p_2 \, ) \,,\nn
&3N:& \quad \langle \vec p_1 \, \vec p_2  \, \vec p_3 | \vec p_1 {}' \,  
\vec p_2 {}'  \,  \vec p_3 {}' \rangle = 
\delta^3 (\vec p_1 {} ' - \vec p_1 \, ) \, \delta^3 (\vec p_2 {}' - \vec p_2
\, ) \,\delta^3 (\vec p_3 {}' - \vec p_3 \, ) \,.
\eeqa
It can be circumvented by assigning a chiral dimension to the transition
operator rather than to its matrix elements in the $N$-nucleon space. 
Adding the factor $3N-6$  to the right-hand side of Eq.~(\ref{pow_orig}) in
order  to account for the normalization 
of the $N$-nucleon states and to ensure that the LO contribution to the nuclear
force appears at order $\nu = 0$ we obtain
\beq
\label{pow_mod}
\nu = -2 + 2 N + 2 (L - C) + \sum_i V_i \Delta_i \,.
\eeq
We will now rewrite this expression in a different form which is better
suited for the method of unitary transformation. To that aim  
we combine Eqs.~(\ref{eq1}) and  (\ref{eq2}) and express $2 ( L - C)$ as
\beq
2 (L - C) =  - 2 N + \sum_i V_i \left( p_i + n_i - 2 \right)\,. 
\eeq
Substituting this expression into  Eq.~(\ref{pow_mod}) leads to
\beq
\label{pow_fin}
\nu = -2 + \sum V_i \kappa_i \,, \quad \quad \kappa_i = d_i + \frac{3}{2} n_i + p_i - 4\,.
\eeq
Clearly, the quantity $\kappa_i$ which enters this expression is just the 
canonical field dimension of a vertex of type $i$ (up to the additional
constant $-4$) and gives the inverse mass dimension of the corresponding
coupling constant. It should further be emphasized that  Eq.~(\ref{pow_fin}) 
can be obtained immediately by counting inverse 
powers of the hard scale $\Lambda$ rather than powers of the soft scale $Q$
(which is, of course, completely  equivalent). Indeed, since the only way 
for the hard scale to be generated is through the values of the low-energy
constants (LECs), the power $\nu$ is just the negative of the overall mass
dimension of all LECs. The
additional factor $-2$ in  Eq.~(\ref{pow_fin}) is a convention 
to ensure that the contributions to the nuclear force start at $\nu = 0$. 
While Eq.~(\ref{pow_fin}) does not say much about the topology and is,
therefore, not particularly useful to
deal with diagrams, it is very convenient for algebraical calculations. 
In particular, it allows to (formally) reduce the chiral expansion 
to the ordinary expansion in
powers of the coupling constant. The role of the coupling constant is played by 
the ratio $Q/\Lambda$, and the power of the coupling constant for a vertex of
type $i$ is given by $\kappa_i$. Clearly, for perturbation theory to be
applicable it is necessary, that only nonrenormalizable (i.e.~the ones with
$\kappa_i > 0$) interactions appear in the Lagrangian. This is guaranteed by the
spontaneously broken chiral symmetry.

\subsection{Application to the method of unitary transformation}
\label{sec:UT}

We are now in the position to apply Eq.~(\ref{pow_fin}) to derive nuclear
forces from the effective chiral Lagrangian using the method of unitary
transformation. The starting point is the time-independent Schr\"odinger equation 
for interacting pions and nucleons
\beq
\label{schroed1}
(H_0 + H_I) | \Psi \rangle = E | \Psi \rangle\,,
\eeq
where $|\Psi \rangle$ denotes an eigenstate of the Hamiltonian $H$ 
with the eigenvalue $E$. Let $\eta$ ($\lambda$)  be 
projection operators onto the purely nucleonic (the remaining) part of the
Fock space satisfying $\eta^2 = \eta$, $\lambda^2 = \lambda$, $ \eta \lambda 
= \lambda \eta = 0$ and $\lambda + \eta = {\bf 1}$. 
To study nuclear systems below the pion production threshold it is advantageous to 
project Eq.~(\ref{schroed1}) onto the $\eta$-subspace of the 
full Fock space. The resulting effective equation can then be solved using 
the standard methods of few- or many-body physics. This reduction can be achieved via
an appropriately chosen unitary transformation 
\beq
\tilde H \equiv U^\dagger H U = \left( \begin{array}{cc} \eta \tilde H \eta  &
    0 
\\ 0 & \lambda \tilde H \lambda \end{array} \right)\,.
\eeq
Following Okubo \citep{Okubo:1954aa}, the unitary operator $U$ can be parametrized as
\begin{equation}
\label{5.9}
U = \left( \begin{array}{cc} \eta (1 +  A^\dagger  A )^{- 1/2} & - 
 A^\dagger ( 1 +  A A^\dagger )^{- 1/2} \\
 A ( 1 +  A^\dagger  A )^{- 1/2} & 
\lambda (1 +  A  A^\dagger )^{- 1/2} \end{array} \right)~,
\end{equation}
with the operator $A= \lambda  A \eta$. The operator $A$ has to satisfy 
the decoupling equation 
\begin{equation}
\label{5.10}
\lambda \left( H - \left[ A, \; H \right] - A H A \right) \eta = 0
\end{equation}
in order for the transformed Hamiltonian $\tilde H$ to be of a block-diagonal form. 
The effective $\eta$-space potential $V$ can be expressed in terms of the operator $A$ as: 
\beq
\label{effpot}
{V} =  \eta (\tilde H  - H_0 )  = \eta \bigg[ (1 + A^\dagger A)^{-1/2} (H + A^\dagger H + H A + A^\dagger H A )  
(1 + A^\dagger A)^{-1/2} - H_0 \bigg] \eta~.
\eeq
The potential $ V$ can be derived perturbatively
based on the chiral power counting. In
Refs.~\cite{Epelbaum:1998ka,Epelbaum:2000kv} this was achieved using
Eq.~(\ref{pow_mod}) via a two-dimensional recursive process which required
rather tedious calculations in the intermediate steps. As an alternative, one
can use Eq.~(\ref{pow_fin}) to express $H_I$ as 
\begin{equation}
\label{n11}
H_I = \sum_{\kappa = 1}^{\infty} H^\kappa
\end{equation}
with $\kappa$ being defined in Eq.~(\ref{pow_fin}). Assuming the following
expansion for the operator $A$
\begin{equation}
\label{n12}
A = \sum_{\kappa = 1}^\infty A^{\kappa}\,,
\end{equation}
one immediately obtains from Eq.~(\ref{5.10}):
\begin{equation}
\label{n13}
A^{\kappa} = \frac{1}{E_\eta - E_\lambda} \lambda \left\{ H^{\kappa} + \sum_{i =
    1}^{\kappa -1} H^i A^{\kappa -i} - \sum_{i=1}^{\kappa -1} A^{\kappa -i} H^i
- \sum_{i = 1}^{\kappa -2} \; \sum_{j =1}^{\kappa - j - 1} A^i H^j A^{\kappa -i-j} 
\right\} \eta
\,. 
\end{equation}  
Here, $E_\eta$ ($E_\lambda$) refers to the free energy of nucleons (nucleons and pions)
in the state $\eta$ ($\lambda$). The expression for the effective potential 
follows immediately by substituting Eqs.~(\ref{n11}) and (\ref{n13}) into
Eq.~(\ref{effpot}).

\section{Derivation of the leading four-nucleon force}
\def\theequation{\arabic{section}.\arabic{equation}}
\label{sec:N3LO}

\subsection{Effective Lagrangian}
\label{sec:el}

The effective chiral Lagrangian for pions and nucleons has the form, see e.g.~\cite{Bernard:1995dp,Bernard:2007zu}:
\beqa
\label{lagr}
\mathcal{L}_{\pi \pi} &=& \frac{F_\pi^2}{4} \, \mbox{tr}  \left[ \partial_\mu U \partial^\mu U^\dagger
+ M_\pi^2 ( U + U^\dagger ) \right] + \ldots \,,\\
\mathcal{L}_{\pi N} &=& N^\dagger \left( i D_0 - \frac{g_A}{2} \vec \sigma
  \cdot \vec u \right) N  + \ldots \,,\nn
\mathcal{L}_{NN} &=& -\frac{1}{2} C_S ( N^\dagger N )  ( N^\dagger N )  - \frac{1}{2} C_T ( N^\dagger \vec \sigma N ) 
\cdot ( N^\dagger \vec \sigma N ) + \ldots \,, 
\nonumber
\eeqa
where only those terms are shown explicitly which contribute to the leading
4NF. Here, $F_\pi = 92.4$ MeV ($g_A = 1.267$) is the 
pion decay constant (the nucleon axial-vector coupling), 
$N$ represents a non-relativistic nucleon field and 
$\vec \sigma$ denote the Pauli spin matrices. The low-energy constants
$C_S$ and $C_T$ determine the strength of the leading NN short-range interaction  
\cite{Weinberg:1990rz,Weinberg:1991um}. Further, the SU(2) matrix $U = u^2$
collects the pion fields, 
$D^\mu = \partial^\mu + \frac{1}{2} [ u^\dagger , \, \partial^\mu u ]$ denotes the covariant derivative of 
the nucleon field and $u_\mu = i u^\dagger \partial_\mu U u^\dagger$. 
The first terms in the expansion of the matrix $U(\fet \pi )$ in powers of the pion fields 
take the form
\beq
\label{alfa_def}
U (\fet \pi ) = 1 + \frac{i}{F_\pi} \fet \tau \cdot \fet \pi - 
\frac{1}{2 F_\pi^2} \fet \pi^2 - \frac{i \alpha}{F_\pi^3} 
(\fet \tau \cdot \fet \pi )^3 + \frac{8 \alpha - 1}{8 F_\pi^4} \fet \pi^4 + \ldots\,,
\eeq
where $\fet \tau$ denote the Pauli isospin matrices and $\alpha$ is an arbitrary constant. 
Notice that only the coefficients in front of the linear and quadratic terms in the pion field are 
fixed uniquely from the unitarity condition $U^\dagger U=1$ and the proper normalization of the 
pion kinetic energy. The explicit $\alpha$-dependence of the matrix $U$
represents the freedom in the definition of the pion field. Clearly, all measurable quantities 
are be $\alpha$-independent. 

Expanding the terms in Eq.~(\ref{lagr}) in powers of pion fields and applying
the canonical formalism leads to
the following interaction terms in the Hamilton density:
\beqa
\label{ham}
\mathcal{H}^{1} &=& \frac{g_A}{2 F_\pi} N^\dagger \fet \tau \vec \sigma
 \cdot \vec \nabla \fet \pi  N \,,\nn
\mathcal{H}^{2} &=& \frac{4 \alpha - 1}{2 F_\pi^2} \,  (\fet \pi \cdot
 \partial_\mu \fet \pi )^2 + \frac{\alpha}{F_\pi^2} \fet \pi^2 (\partial_\mu
 \fet \pi \cdot \partial^\mu \fet \pi ) - \frac{8 \alpha - 1}{8 F_\pi^2}
 M_\pi^2 \fet \pi^4
+ \frac{1}{4 F_\pi^2} N^\dagger \fet \tau \cdot (\fet \pi
 \times \dot{\fet \pi} ) N \nn
&& {} + \frac{1}{2} \, C_S ( N^\dagger N )  ( N^\dagger N ) 
+ \frac{1}{2} C_T \, ( N^\dagger \vec \sigma N ) \cdot ( N^\dagger \vec \sigma N ) \,, \nn
\mathcal{H}^{3} &=& - \frac{g_A}{2 F_\pi^3} N^\dagger \left[ \left( 2 \alpha -
 \frac{1}{2} \right) (\fet \tau \cdot \fet \pi ) ( \fet \pi \vec \sigma \cdot
 \vec \nabla \fet \pi ) + \alpha \fet \pi^2 (\fet \tau \vec \sigma \cdot \vec
 \nabla \fet \pi ) \right] N \,, \nn
\mathcal{H}^{4} &=& \frac{1}{2 (2  F_\pi )^2}  [ N^\dagger (\fet \tau \times \fet \pi ) N ]
 \cdot   [ N^\dagger (\fet \tau \times \fet \pi ) N ] \,.
\eeqa
Here, the superscripts of $\mathcal{H}$ refer to the canonical field dimension
 $\kappa$ defined in Eq.~(\ref{pow_fin}).
Again, we only show those terms in the Hamilton density which are relevant 
for the present calculation. Notice that the last term in Eq.~(\ref{ham}) is
 absent in the effective Lagrangian in Eq.~(\ref{lagr}) and 
arises  through the application of the canonical formalism, see 
e.g.~\cite{Weinberg:1992yk}. 

\subsection{Orders $\nu < 4$.}
\label{sec:ord2}

\begin{figure}[tb]
\vskip 1 true cm
\includegraphics[width=4.5cm,keepaspectratio,angle=0,clip]{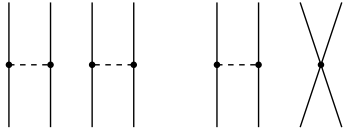}
    \caption{
         Disconnected 4N diagrams at order $\nu = 2$. Solid and dashed lines
         represent nucleons and pions,
         respectively. Solid dots denote vertices with $\Delta_i = 0$ from Eq.~(\ref{ham}).
\label{fig1} 
 }
\end{figure}

\begin{figure}[tb]
\vskip 1 true cm
\includegraphics[width=10.5cm,keepaspectratio,angle=0,clip]{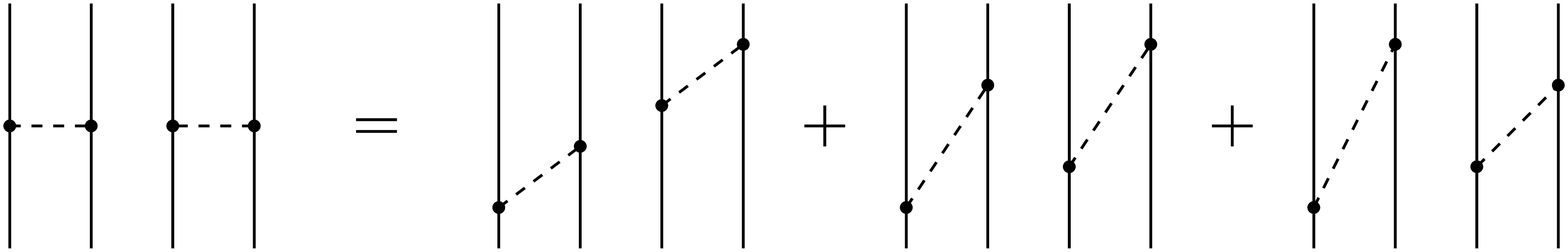}
    \caption{
         Different ``time orderings'' for the first diagram in Fig.~\ref{fig1}.
         respectively. Graphs resulting by the interchange of the nucleon
         lines are not shown. 
\label{fig1a} 
 }
\end{figure}

The structure of the effective Hamilton operator at leading order can be worked out
straightforwardly applying the projection formalism as described in section 
\ref{sec:UT} to the lowest-order (i.e.~with $\Delta_i = 0$) Hamilton density in
Eq.~(\ref{ham}). This leads to 
\beq
\label{LO}
V^{(0)}= \eta \Big[ H_{40}^2 - H_{21}^1 \frac{\lambda^1}{E_\pi} H_{21}^1
\Big] \eta\,.
\eeq
Here, the superscript of $\lambda$ refers to the number of pions in the 
corresponding intermediate state, $E_\pi$ is the total energy
of $n$ pions in the state $\lambda^n$, $E_\pi = \sum_{i=1}^n \omega_i$
where  $\omega_i = \sqrt{\vec q_i \, ^2 + M_\pi^2}$
is the energy of the pion with momentum $\vec q_i$. The
subscripts $a$ and $b$ of $H_{ab}^{\kappa}$ denote the number of nucleon and
pion field operators, respectively. They are introduced in order to clarify
the meaning of various terms in the effective potential. 
Clearly, the terms in Eq.~(\ref{LO}) only give rise to one- and two-nucleon
operators. The first corrections to the effective Hamiltonian arise at
order $\nu = 2$. The relevant terms at this order read:
\beqa
\label{NLO}
V^{(2)} &=& \eta \Big[ 
- H_{21}^1 \frac{\lambda^1}{E_\pi} H_{21}^1 \frac{\lambda^2}{E_\pi} H_{21}^1 
\frac{\lambda^1}{E_\pi} H_{21}^1 + \frac{1}{2} H_{21}^1
\frac{\lambda^1}{E_\pi^2} H_{21}^1 \, \eta \,
H_{21}^1 \frac{\lambda^1}{E_\pi} H_{21}^1 + \frac{1}{2} H_{21}^1 \frac{\lambda^1}
{E_\pi} H_{21}^1 \, \eta \,
H_{21}^1 \frac{\lambda^1}{E_\pi^2} H_{21}^1   \nn
&& {} 
+ H_{21}^1  \frac{\lambda^1}{E_\pi} 
H_{40}^2 \frac{\lambda^1}{E_\pi} H_{21}^1 - \frac{1}{2}
H_{21}^1 \frac{\lambda^1}{E_\pi^2} H_{21}^1 \, \eta \, H_{40}^2  - \frac{1}{2}
H_{40}^2 \, \eta \,  H_{21}^1  
\frac{\lambda^1}{E_\pi^2} H_{21}^1  
\Big] \eta\,.
\eeqa
Here we only list those terms which can generate four-nucleon operators.
The complete list of terms at this order can be found in 
\cite{Epelbaum:1998ka,Epelbaum:2002gb}. The contribution to the 4NF from the operators in 
Eq.~(\ref{NLO}) can be represented schematically by disconnected diagrams
shown in Fig.~\ref{fig1}. 
It should be understood that diagrams in the method of unitary transformation
have a different meaning compared to the ones arising in the context of time-ordered
perturbation theory and serve merely to visualize the topology corresponding
to a given sequence of operators $H_{ab}^\kappa$. We further emphasize that 
diagrams shown in this work represent the sum over all possible ``time orderings''
as depicted in Fig.~\ref{fig1a}. 

The contributions to the 4NF from diagrams in Fig.~\ref{fig1} have been
considered in Ref.~\cite{vanKolck:1994yi} using time-ordered perturbation
theory. In this approach, only the first and the fourth terms in
Eq.~(\ref{NLO}) contribute to the effective Hamiltonian.  Evaluating matrix
elements of these terms corresponding to diagrams shown in Fig.~\ref{fig1a}
one obtains non-vanishing contributions to the 4NF. As pointed out in
Ref.~\cite{vanKolck:1994yi}, the resulting 4NF cancels
against the
recoil corrections to the 2NF upon the iteration in the dynamical
equation. The same sort of cancellation also takes place for the 3NF at
$\nu = 2$ \cite{Yang:1986pk,vanKolck:1994yi}. 

In the method of unitary transformation, one has to take into account all
terms in Eq.~(\ref{NLO}). It is easy to verify that the resulting 4NF vanishes 
due to an exact cancellation between the different terms 
in this equation. A similar cancellation occurs also for the 3NF
\cite{Eden:1996ey,Epelbaum:2000kv}, see also
\cite{Coon:1986kq}, and for disconnected 2NF
diagrams \cite{Epelbaum:2002gb} at this chiral order. 
The corrections to the effective Hamilton operator at order $\nu = 3$ within
the method of unitary transformation are discussed in detail in
\cite{Epelbaum:2000kv}. They give rise to at most tree-nucleon operators
and will not be discussed in this work.

\subsection{Order $\nu = 4$.}
\label{sec:ambig}

The first non-vanishing contributions to the 4NF arise at order $\nu =
4$. For the sake of a better overview, it is useful to divide the total
contribution at this order into pieces with the same combinations of the coupling
constants $g_A$ and $C_{S,T}$. This leads to eight classes of terms which are
discussed in detail below.  

\begin{itemize}
\item
{\bf Class-I} contributions proportional to $g_A^6$.

This class of contributions arises from all possible 4NF diagrams
involving six vertices $H_{21}^1$ from the first line of
Eq.~(\ref{ham}). Using Eqs.~(\ref{effpot})-(\ref{n13}) one obtains:
\beqa
\label{class1}
V^{(4)}&=& \frac{1}{2} \eta \bigg[ H_{21}^1 
\frac{\lambda^1}{E_\pi}  H_{21}^1 \, \eta \, H_{21}^1
  \frac{\lambda^1}{E_\pi}  H_{21}^1  \frac{\lambda^2}{E_\pi}
   H_{21}^1 \frac{\lambda^1}{E_\pi^2} \, H_{21}^1 \;
+ \; H_{21}^1 \frac{\lambda^1}{E_\pi} H_{21}^1 \, \eta \, H_{21}^1
  \frac{\lambda^1}{E_\pi} H_{21}^1 \frac{\lambda^2}{E_\pi^2}
  H_{21}^1 \frac{\lambda^1}{E_\pi} H_{21}^1 \nn [2pt]
&& {}
+ H_{21}^1 \frac{\lambda^1}{E_\pi} H_{21}^1 \, \eta \, H_{21}^1
  \frac{\lambda^1}{E_\pi^2} H_{21}^1 \frac{\lambda^2}{E_\pi}
  H_{21}^1 \frac{\lambda^1}{E_\pi} H_{21}^1 
\; + \;  H_{21}^1 \frac{\lambda^1}{E_\pi^2} H_{21}^1 \, \eta \, H_{21}^1
  \frac{\lambda^1}{E_\pi} H_{21}^1 \frac{\lambda^2}{E_\pi}
  H_{21}^1 \frac{\lambda^1}{E_\pi} H_{21}^1 \nn [2pt]
&& {}
- H_{21}^1 \frac{\lambda^1}{E_\pi} H_{21}^1 \, \eta \, H_{21}^1
  \frac{\lambda^1}{E_\pi} H_{21}^1 \,  \eta \,
  H_{21}^1 \frac{\lambda^1}{E_\pi^3} H_{21}^1 
\; - \;\frac{1}{4} H_{21}^1 \frac{\lambda^1}{E_\pi^2} H_{21}^1 \, \eta \, H_{21}^1
  \frac{\lambda^1}{E_\pi} H_{21}^1 \,  \eta \,
  H_{21}^1 \frac{\lambda^1}{E_\pi^2} H_{21}^1 \nn [2pt]
&& {}
- \frac{3}{4} H_{21}^1 \frac{\lambda^1}{E_\pi} H_{21}^1 \, \eta \, H_{21}^1
  \frac{\lambda^1}{E_\pi^2} H_{21}^1 \,  \eta \,
  H_{21}^1 \frac{\lambda^1}{E_\pi^2} H_{21}^1 \; - \;  
  H_{21}^1 \frac{\lambda^1}{E_\pi} H_{21}^1 \frac{\lambda^2}{E_\pi
  } H_{21}^1 \frac{\lambda^1}{E_\pi} H_{21}^1 \frac{\lambda^2}{E_\pi
  } H_{21}^1 \frac{\lambda^1}{E_\pi} H_{21}^1 \nn [2pt]
&& {}
  - H_{21}^1 \frac{\lambda^1}{E_\pi} H_{21}^1 \frac{\lambda^2}{E_\pi
  } H_{21}^1 \frac{\lambda^3}{E_\pi} H_{21}^1 
  \frac{\lambda^2}{E_\pi} H_{21}^1 \frac{\lambda^1}{E_\pi} H_{21}^1  \bigg] \eta + \mbox{h.c.}\,.
\eeqa
The corresponding contributions to the 4NF can, in principle, be evaluated 
straightforwardly by calculating matrix elements of the operators in
Eq.~(\ref{class1}) for all possible ``time orderings'' of diagrams shown in 
Fig.~\ref{fig3}. 
\begin{figure}[tb]
\vskip 1 true cm
\includegraphics[width=16.0cm,keepaspectratio,angle=0,clip]{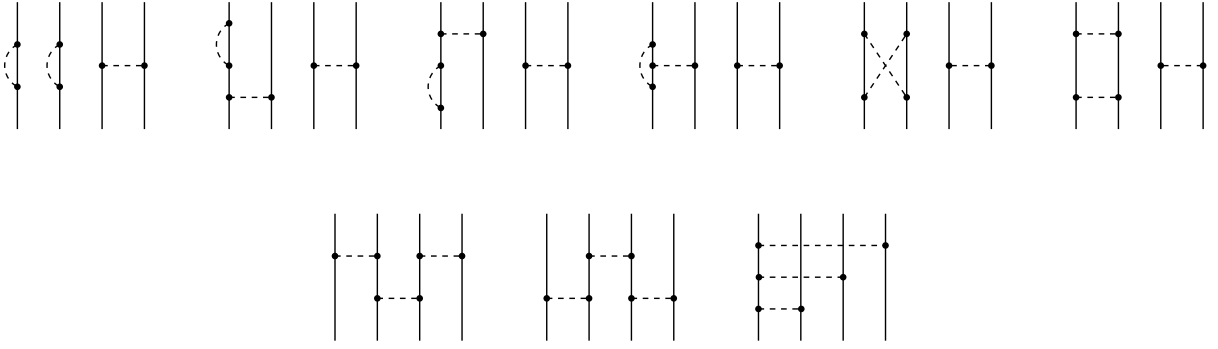}
    \caption{
         Class-I contributions to the 4NF. For 
         notation, see Figs.~\ref{fig1} and \ref{fig1a}.
\label{fig3} 
 }
\end{figure}
Before giving the explicit results, it is important to address the issue whether the
contribution $V^{(4)}$ is defined unambiguously.    
Generally, an effective Hamilton operator is defined modulo unitary 
transformations.\footnote{Other sources of ambiguities related to field
redefinitions in the Lagrangian and to the choice of the dynamical equation
are discussed for one- and two-pion exchange potentials in Ref.~\cite{Friar:1999sj}.} 
It should, therefore, be understood that terms in
Eq.~(\ref{class1}) correspond to \emph{one}
particular choice of the unitary transformation $U$. Indeed, Eq.~(\ref{5.9}) 
does not yield the most general parametrization of the operator $U$. 
The resulting effective Hamilton operator $\eta \tilde H \eta$ can be
further modified via subsequent unitary transformations acting on the $\eta$-space.  
To be specific, consider the unitary transformation of the
form:
\beq
\label{UT}
U = e^S\,,
\eeq
where $S$ is an anti-hermitian operator acting on the $\eta$-space, $S \equiv
\eta S \eta$, $S^\dagger = - S$. Further, let $S$ be given by $S = \alpha_1
S_1 + \alpha_2 S_2$ with $\alpha_{1,2}$ being arbitrary real numbers and 
\beqa
\label{add1}
S_1 &=& \eta  \bigg[ H_{21}^1 
\frac{\lambda^1}{E_\pi}  H_{21}^1 \, \eta \, H_{21}^1
  \frac{\lambda^1}{E_\pi^3}  H_{21}^1  \; - \;  H_{21}^1 
\frac{\lambda^1}{E_\pi^3}  H_{21}^1 \, \eta \, H_{21}^1
  \frac{\lambda^1}{E_\pi}  H_{21}^1  \bigg] \eta \,, \nn
S_2 &=& \eta \bigg[   H_{21}^1 
\frac{\lambda^1}{E_\pi}  H_{21}^1 \frac{\lambda^2}{E_\pi} H_{21}^1
  \frac{\lambda^1}{E_\pi^2}  H_{21}^1   -   H_{21}^1 
\frac{\lambda^1}{E_\pi^2}  H_{21}^1 \frac{\lambda^2}{E_\pi}   H_{21}^1
  \frac{\lambda^1}{E_\pi}  H_{21}^1 \bigg] \eta \,.
\eeqa
The operators $S_1$ and $S_2$ are the only possible ones that are invariant
under the time-reversal operation and can be
constructed out of four vertices $H_{21}^1$ provided the $E_\pi$'s are only
allowed to appear in the denominators. 
Acting with the transformation $U$ onto the
lowest-order effective Hamilton operator, 
\beq
\label{H0}
H^{(0)} = \eta \bigg[ H_{20}^2 + H_{40}^2 - H_{21}^1 
\frac{\lambda^1}{E_\pi}  H_{21}^1 \bigg] \eta\,, 
\eeq
where $H_{20}^2$ is the nonrelativistic kinetic energy term,
the following additional terms in $V^{(4)}$ are generated:
\beqa
\label{class1_add}
\delta  V^{(4)} &=& [ H^{(0)}, \; S ] \nn
&=& - \alpha_1 \, \eta \bigg[  
 H_{21}^1 \frac{\lambda^1}{E_\pi}  H_{21}^1 \, \eta \,  H_{21}^1
 \frac{\lambda^1}{E_\pi}  H_{21}^1 \,  \eta \, H_{21}^1
  \frac{\lambda^1}{E_\pi^3}  H_{21}^1 \; - \;  
H_{21}^1 \frac{\lambda^1}{E_\pi}  H_{21}^1  \, \eta \, H_{21}^1 
\frac{\lambda^1}{E_\pi^3}  H_{21}^1 \, \eta \, H_{21}^1
  \frac{\lambda^1}{E_\pi}  H_{21}^1 \bigg] \eta \nn
&& {}- \alpha_2 \, \eta \bigg[  H_{21}^1 \frac{\lambda^1}{E_\pi}  H_{21}^1 \, \eta \,  
 H_{21}^1 
\frac{\lambda^1}{E_\pi}  H_{21}^1 \frac{\lambda^2}{E_\pi} H_{21}^1
  \frac{\lambda^1}{E_\pi^2}  H_{21}^1 \; - \; 
H_{21}^1 \frac{\lambda^1}{E_\pi}  H_{21}^1 \, \eta \,  
H_{21}^1 
\frac{\lambda^1}{E_\pi^2}  H_{21}^1 \frac{\lambda^2}{E_\pi}   H_{21}^1
  \frac{\lambda^1}{E_\pi}  H_{21}^1 \bigg] \eta \nn
&& {} + \; \mbox{h.c.} \; + \; \ldots\,.
\eeqa
Here, the ellipses refer to terms involving $H_{20}^2$ ($H_{40}^2$) which give rise to
class-VIII (class-IV) corrections and will
be considered below.   
The class-I contributions to the effective Hamiltonian, therefore,
seem to be defined modulo the arbitrary constants $\alpha_1$ and
$\alpha_2$. We will now demonstrate that there exist one particular
choice for these constants which is strongly preferable. 
To that aim consider \emph{three-nucleon} forces generated by
terms in Eqs.~(\ref{class1}), (\ref{class1_add}) and involving one-pion
exchange between two nucleons, see Fig.~\ref{fig4a}, diagrams (a)-(f). We can 
generally express the structure of the 3NF in the form 
\beq
V^i_{3N} = M^i_{3N} \, \bar V^i_{3N}\,.
\eeq
\begin{figure}[tb]
\vskip 1 true cm
\includegraphics[width=16.0cm,keepaspectratio,angle=0,clip]{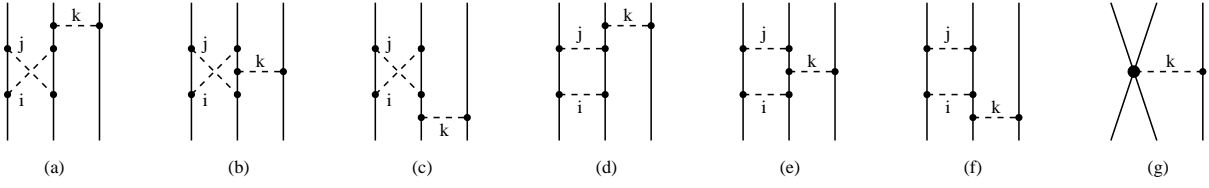}
    \caption{
         Selected three-pion exchange contributions to the
         3NF at order $\nu = 4$ (graphs (a)-(f)) and one-pion exchange 3NF at
         order $\nu =4$. Filled circle denotes the leading $\pi NNNN$ vertex
         with $\Delta_i =1$.  For remaining 
         notation, see Figs.~\ref{fig1} and \ref{fig1a}.
\label{fig4a} 
 }
\end{figure}
Here, $M^i_{3N}$ represents the spin, isospin and momentum structure which results
from vertices entering a diagram $i$ and $\bar V^i_{3N}$ denotes 
the corresponding energy denominators. 
Evaluating matrix elements of operators in Eqs.~(\ref{class1}) and
(\ref{class1_add}) for all possible ``time orderings'' of diagrams (a)-(f) in
Fig.~\ref{fig4a} yields
\beqa
\label{match_class1}
\bar V^{\rm a}_{3N} &=&  \frac{4}{\omega_i^3 \omega_j \omega_k} \,, \nn
\bar V^{\rm b}_{3N} &=& - \frac{4}{\omega_i^3 \omega_j \omega_k} -
\frac{4}{\omega_i \omega_j^3 \omega_k}  \,, \nn
\bar V^{\rm c}_{3N} &=& \bar V^{\rm a}_{3N}\, \Big|_{\{a,i,j\} \to \{c,j,i\}}  \,, \nn
\bar V^{\rm d}_{3N} &=&  - 4 \frac{1 - 2 \alpha_2 }{\omega_i \omega_j \omega_k^3}
+ \frac{1 + 2 \alpha_1}{\omega_i \omega_j^2 \omega_k^2}
+ 2 \frac{1 + 2 \alpha_1 }{\omega_i \omega_j^3 (\omega_j + \omega_k )} \nn
&& {} + \frac{1}{\omega_k} \bigg( - 4 \frac{1 -
  2 \alpha_2 }{\omega_i^3
  \omega_j} + 2 \frac{1 - 2 \alpha_1 - 8 \alpha_2}{\omega_i \omega_j^3} - 
\frac{1 + 2 \alpha_1}{\omega_i \omega_j^2 (\omega_i + \omega_j ) }
+ \frac{1 + 2 \alpha_1}{\omega_i^2 \omega_j (\omega_i + \omega_j)} \bigg)
\,, \nn
\bar V^{\rm e}_{3N} &=&   8 \frac{1 - 2 \alpha_2 }{\omega_i \omega_j \omega_k^3}
- \frac{1 + 2 \alpha_1}{\omega_i \omega_j^2 \omega_k^2}
- \frac{1 + 2 \alpha_1}{\omega_i^2 \omega_j \omega_k^2}
- 2 \frac{1 + 2 \alpha_1 }{\omega_i \omega_j^3 (\omega_j + \omega_k )} 
- 2 \frac{1 + 2 \alpha_1 }{\omega_i^3 \omega_j (\omega_i + \omega_k )} 
\nn 
&& {} + \frac{1}{\omega_k} \bigg( 
2 \frac{1 +2 \alpha_1 + 4 \alpha_2}{\omega_i \omega_j^3}
+ 2 \frac{1 +2 \alpha_1 + 4 \alpha_2}{\omega_i^3 \omega_j} \bigg)  \,, \nn
\bar V^{\rm f}_{3N} &=& \bar V^{\rm d}_{3N}\, \Big|_{\{d,i,j\} \to \{f,j,i\}}\,.
\eeqa
The contributions from crossed-box diagrams (a-c) exhibit the
expected dependence on the pion energy $\omega_k$ which leads to the usual 
static one-pion exchange potential (OPEP) $\propto \omega_k^{-2} = 1/(\vec q_k\, ^2 +
M_\pi^2 )$. The additional factor of $\omega_k^{-1}$ results from the pion
phase-space factors. Notice further that $V^{\rm a,b,c}_{3N}$ do not depend 
on the parameters $\alpha_{1,2}$\footnote{The independence on $\alpha_1$
  follows directly from Eq.~(\ref{class1_add}).}. The contributions from the box
diagrams (d-f) do, however, depend on $\alpha_{1,2}$ and involve terms that
cannot be associated with the static OPEP between the second and the
third nucleons. For example, terms proportional to $\omega_k^{-2}$ in the last
three lines of Eq.~(\ref{match_class1}) lead to the 3N potential  
 proportional to $(\vec q_k\, ^2 + M_\pi^2 )^{-3/2}$ which has a cut 
starting at $\vec q \, ^2 = - M_\pi^2$. The presence of such a cut 
which is not produced by multi-particle intermediate states appears, at first
sight, rather puzzling. One should, however, keep in mind that we are dealing
here with the non-iterative part of the scattering amplitude 
which gives rise to nuclear forces. Its analytic
structure needs, in principle, not to coincide with the one of the  
full amplitude. A more serious problem due to the appearance of the ``unphysical''
terms in Eq.~(\ref{match_class1}) is non-renormalizability of the
corresponding 3NF. Indeed, one can easily verify using dimensional arguments
that the loop integrals entering, for example, the $\omega_k^{-3}$
($\omega_k^{-2}$) terms  give rise to cubic and linear (quadratic and
logarithmic) ultraviolet divergences. The only relevant counter term which is
available at this order and which can be used to absorb the infinities is 
$H_{41}^4$ with one pion and four nucleon field operators. It is proportional
to the LEC $D$ and enters the 3NF at order $\nu = 3$, see diagram (g)
in Fig.~\ref{fig4a}, which arises from
\beq
\label{3NF_NNLO}
V^{(3)}= - \eta \Big[ H_{41}^4 \frac{\lambda^1}{\omega_k} H_{21}^1 + 
 H_{21}^1 \frac{\lambda^1}{\omega_k} H_{41}^4
\Big] \eta\,,
\eeq
and leads to $\bar V^{\rm g}_{3N} = -2/\omega_k$ \cite{vanKolck:1994yi,Epelbaum:2002vt}. The
 contributions to the 3NF in Eq.~(\ref{match_class1}) with a different
 dependence on $\omega_k$ can, obviously, not be renormalized in this way. Again, it
 should be understood that
the above mentioned difficulty does not affect S-matrix elements. All 
``problematic'' divergences that enter the 3NF have to cancel against the
divergences arising from iterative contributions to the amplitude in such
a way that the resulting
S-matrix is renormalizable in the usual sense. For the purpose
of describing the few-nucleon dynamics based on the Schr\"odinger equation, it is,
however, desirable to have  a well-defined effective Hamilton operator.   
The difficulty with the non-renormalizability of the 3NF in
Eq.~(\ref{match_class1}) can be avoided if one requires that only the
$\omega_k^{-1}$ terms are present, i.e. the OPEP factorizes out in  
diagrams (d-f) in Fig.~\ref{fig4a}. This can be achieved via a suitable choice
 of the $\eta$-space unitary transformation by setting  
\beq
\label{alpha_1_2}
\alpha_1 =  - \alpha_2 = -\frac{1}{2}\,.
\eeq
This choice leads to the following remarkably simple expressions:
\beqa
\label{class1_box}
\bar V^{\rm d}_{3N} &=& -  \frac{4}{\omega_i \omega_j^3 \omega_k} \,, \nn
\bar V^{\rm e}_{3N} &=&   \frac{4}{\omega_i^3 \omega_j \omega_k} +
\frac{4}{\omega_i \omega_j^3 \omega_k}  \,, \nn
\bar V^{\rm f}_{3N} &=& V^{\rm d}_{3N}\, \Big|_{\{d,i,j\} \to \{f,j,i\}}  \,.
\eeqa
These considerations may remind one of the recent 
findings in the context of large-$N_c$ QCD 
\cite{Belitsky:2002ni,Cohen:2002qn,Cohen:2002im}. There it was
found that the multiple-meson exchange potential derived in the energy-dependent
formulation is inconsistent with large-$N_C$ counting rules. The consistency
could be maintained using a different (but completely equivalent)
form of the potential based on the energy-independent formalism, 
see Ref.~\cite{Cohen:2002im} for more details.

Choosing the parameters $\alpha_{1,2}$ as described above, it is a
straightforward exercise to calculate the contributions to the 4NF associated
with the diagrams in Fig.~\ref{fig3}. We found that all disconnected graphs
lead to vanishing contributions regardless of the values of 
$\alpha_{1,2}$. From the connected diagrams in the second row in
Fig.~\ref{fig3}, only the first two generate non-vanishing contributions to
the 4NF which take the form:
\beqa
\label{4nf_class1}
V_{\rm Class-I} &=& - \frac{2 g_A^6}{( 2 F_\pi )^6} 
\frac{\vec \sigma_1 \cdot \vec q_1 \;\vec \sigma_4 \cdot \vec q_4}{[\vec q_1^{\;2}  + M_\pi^2]\,
[\vec q_{12}^{\;2}  + M_\pi^2]^2 \, [\vec q_4^{\;2}  + M_\pi^2]} \nn
&\times & {} \Big[ ( \fet \tau_1 \cdot \fet \tau_4 \,  \fet \tau_2 \cdot \fet \tau_3
-  \fet \tau_1 \cdot \fet \tau_3 \,  \fet \tau_2 \cdot \fet \tau_4 ) \,\vec q_1 \cdot \vec q_{12}
\, \vec q_4 \cdot \vec q_{12} 
 + \fet \tau_1 \times \fet \tau_2 \cdot \fet \tau_4  \; \vec q_1 \cdot \vec q_{12}  \; 
\vec q_{12} \times \vec q_4 \cdot \vec \sigma_3 \nn
&& {}  + \fet \tau_1 \times \fet \tau_3 \cdot \fet \tau_4  \; \vec q_4 \cdot \vec q_{12}  \; 
\vec q_{1} \times \vec q_{12} \cdot \vec \sigma_2 
 + \fet \tau_1 \cdot \fet \tau_4 \; \vec q_{12} \times \vec q_{1} \cdot \vec \sigma_2 \;
\vec q_{12} \times \vec q_4 \cdot \vec \sigma_3  \Big] +\mbox{all perm.}.
\eeqa
Here, the subscripts refer to the nucleon labels and $\vec q_{i} = \vec p_i \, ' - \vec p_i$ with $\vec p_i \, '$
and $\vec p_i$ being the final and initial momenta of the nucleon $i$. 
Further, $\vec q_{12} = \vec q_1 + \vec q_2 = - \vec q_3 - \vec q_4 = -\vec
q_{34}$ is the momentum transfer between the 
nucleon pairs 12 and 34.
We have verified that the obtained expression for the class-I 4NF remains
unchanged if one considers a larger class of unitary transformations in Eq.~(\ref{UT})
with the operator $S$ involving terms with pion energies in the numerators. 
Although the renormalizability condition does not completely fix the
corresponding $\eta$-space unitary transformations in that case, the remaining
ambiguity does not affect nuclear forces at the considered order. 
Notice further that the contribution from the last
graph in Fig.~\ref{fig3} only vanishes if the constants $\alpha_{1,2}$ are chosen
according to Eq.~(\ref{alpha_1_2}).
Finally, it should be emphasized that the ambiguity due to the
additional $\eta$-space unitary transformations does not show up at lower orders
in the chiral expansion since it is not possible to construct non-vanishing
anti-hermitian operators $S$ with just two vertices $H_{21}^1$.

\item
{\bf Class-II} contributions proportional to $g_A^4$.

These contributions arise from 4N diagrams
involving four vertices $H_{21}^1$ and one insertion of the 
Weinberg-Tomozawa vertex $H_{22}^2$, see Fig.~\ref{fig5}. In addition, there
are diagrams with one insertion of the nonlinear pion and
pion-nucleon interactions $H_{04}$ and $H_{23}$, respectively. 
Using Eqs.~(\ref{effpot})-(\ref{n13}) one obtains the following contributions
to the class-II effective Hamilton operator:
\beqa
\label{class2}
V^{(4)} &=&  \eta \bigg[ - \frac{1}{2}  
  H_{21}^1 \frac{\lambda^1}{E_\pi} H_{21}^1 \, \eta \, H_{21}^1 \frac{\lambda^1}{E_\pi} H_{21}^1
  \frac{\lambda^2}{E_\pi^2} H_{22}^2  
\; - \; \frac{1}{2}  
  H_{21}^1 \frac{\lambda^1}{E_\pi} H_{21}^1 \, \eta \, H_{21}^1 \frac{\lambda^1}{E_\pi} H_{22}^2
  \frac{\lambda^1}{ E_\pi^2} H_{21}^1 
  \nn[2pt]
&& {}
- \frac{1}{2}  
  H_{21}^1 \frac{\lambda^1}{E_\pi} H_{21}^1 \, \eta \, H_{22}^2 \frac{\lambda^2}{E_\pi
  } H_{21}^1  \frac{\lambda^1}{ E_\pi^2} H_{21}^1 
\;   - \; \frac{1}{2}  
  H_{21}^1 \frac{\lambda^1}{E_\pi} H_{21}^1 \, \eta \, H_{21}^1 \frac{\lambda^1}{E_\pi^2} H_{21}^1
  \frac{\lambda^2}{ E_\pi} H_{22}^2   \nn [2pt]
&& {}
- \frac{1}{2}  
  H_{21}^1 \frac{\lambda^1}{E_\pi} H_{21}^1 \, \eta \, H_{21}^1 \frac{\lambda^1}{E_\pi^2} H_{22}^2
  \frac{\lambda^1}{ E_\pi} H_{21}^1 
\; - \; \frac{1}{2}  
  H_{21}^1 \frac{\lambda^1}{E_\pi} H_{21}^1 \, \eta \, H_{22}^2 \frac{\lambda^2}{E_\pi
  ^2} H_{21}^1  \frac{\lambda^1}{ E_\pi} H_{21}^1   \nn [2pt]
&& {}
- \frac{1}{2}  
  H_{21}^1 \frac{\lambda^1}{E_\pi^2} H_{21}^1 \, \eta \, H_{21}^1 \frac{\lambda^1}{E_\pi} H_{21}^1
  \frac{\lambda^2}{ E_\pi} H_{22}^2  
\; - \; \frac{1}{2}  
  H_{21}^1 \frac{\lambda^1}{E_\pi^2} H_{21}^1 \, \eta \, H_{21}^1 \frac{\lambda^1}{E_\pi} H_{22}^2
  \frac{\lambda^1}{ E_\pi} H_{21}^1   \nn [2pt]
&& {}
- \frac{1}{2}  
  H_{21}^1 \frac{\lambda^1}{E_\pi^2} H_{21}^1 \, \eta \, H_{22}^2 \frac{\lambda^2}{E_\pi
  } H_{21}^1  \frac{\lambda^1}{ E_\pi} H_{21}^1 
\; + \; \frac{1}{2} H_{21}^1 \frac{\lambda^1}{E_\pi} H_{21}^1 \frac{\lambda^2}{E_\pi } H_{22}^2
\frac{\lambda^2}{E_\pi} H_{21}^1
  \frac{\lambda^1}{ E_\pi} H_{21}^1 \nn [2pt]
&& {}
+ H_{21}^1 \frac{\lambda^1}{E_\pi} H_{21}^1 \frac{\lambda^2}{E_\pi } H_{21}^1
\frac{\lambda^1}{E_\pi} H_{21}^1  \frac{\lambda^2}{E_\pi} H_{22}^2
\; + \; H_{21}^1 \frac{\lambda^1}{E_\pi} H_{21}^1 \frac{\lambda^2}{E_\pi } H_{21}^1
\frac{\lambda^1}{E_\pi} H_{22}^2  \frac{\lambda^1}{E_\pi} H_{21}^1 \nn [2pt]
&& {}
+ H_{21}^1 \frac{\lambda^1}{E_\pi} H_{21}^1 \frac{\lambda^2}{E_\pi } H_{21}^1
\frac{\lambda^3}{E_\pi} H_{21}^1
\frac{\lambda^2}{E_\pi} H_{22}^2 \;
+ \; H_{21}^1 \frac{\lambda^1}{E_\pi} H_{21}^1 \frac{\lambda^2}{E_\pi } H_{21}^1
\frac{\lambda^3}{E_\pi} H_{22}^2  \frac{\lambda^1}{E_\pi} H_{21}^1\nn  [2pt]
&& {} 
 -  H_{21}^1 \frac{\lambda^1}{E_\pi} H_{21}^1 \frac{\lambda^2}{E_\pi } H_{21}^1
\frac{\lambda^3}{E_\pi} H_{23}^3 \;
- \; H_{21}^1 \frac{\lambda^1}{E_\pi} H_{21}^1 \frac{\lambda^2}{E_\pi} H_{23}^3
\frac{\lambda^1}{E_\pi} H_{21}^1 \nn  [2pt]
&& {}
+   H_{21}^1 \frac{\lambda^1}{E_\pi} H_{21}^1 \frac{\lambda^2}{E_\pi} H_{21}^1
\frac{\lambda^3}{E_\pi} H_{21}^1
\frac{\lambda^4}{E_\pi} H_{04}^2 \; 
+  \; H_{21}^1 \frac{\lambda^1}{E_\pi} H_{21}^1 \frac{\lambda^2}{E_\pi } H_{21}^1
\frac{\lambda^3}{E_\pi} H_{04}^2 
\frac{\lambda^1}{E_\pi} H_{21}^1  \\  [2pt]
&& {} 
+ \frac{1}{2} H_{21}^1 \frac{\lambda^1}{E_\pi} H_{21}^1 \frac{\lambda^2}{E_\pi} H_{04}^2
\frac{\lambda^2}{E_\pi} H_{21}^1 
\frac{\lambda^1}{E_\pi} H_{21}^1 
\bigg ] \eta +  \mbox{h.c.}\,. \nonumber 
\eeqa
\begin{figure}[tb]
\vskip 1 true cm
\includegraphics[width=16.0cm,keepaspectratio,angle=0,clip]{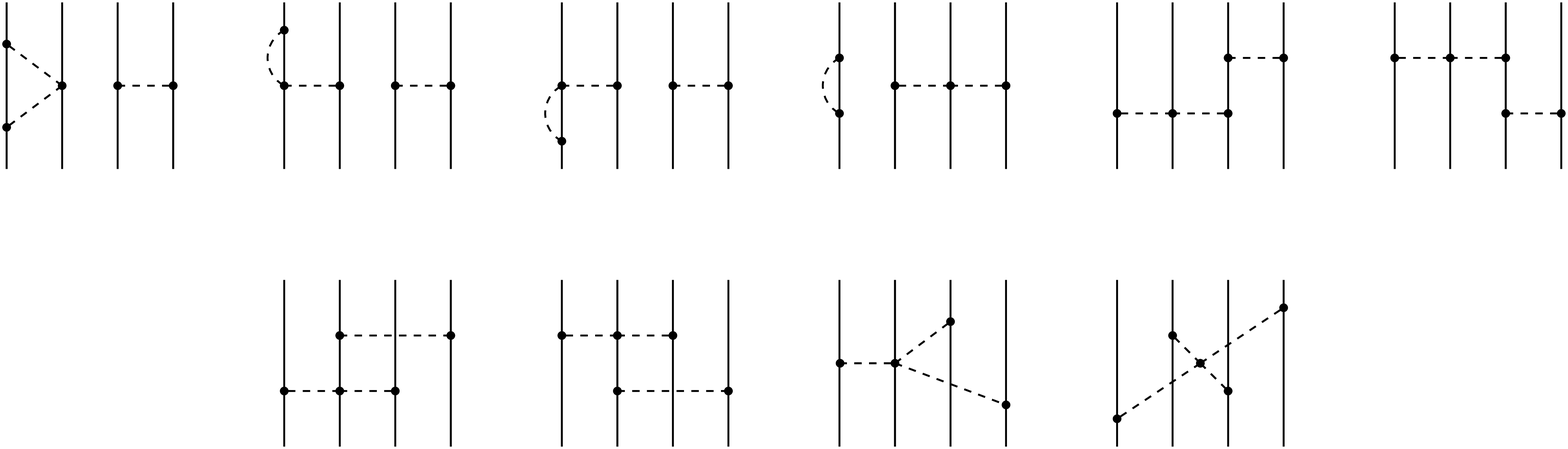}
    \caption{
         Class-II contributions to the 4NF. For 
         notation, see Figs.~\ref{fig1} and \ref{fig1a}.
\label{fig5} 
 }
\end{figure}
Similar to the previously considered class-I contributions, we employ
additional $\eta$-space unitary transformations with the operator $S$ in Eq.~(\ref{UT}) 
given by $S = \alpha_3 S_3 + \alpha_4 S_4 + \alpha_5 S_5$ where $\alpha_i$ are
real constants and
\beqa
\label{add2}
S_3 &=& \eta  \bigg[ H_{21}^1 
\frac{\lambda^1}{E_\pi^2}  H_{22}^2 \, \frac{\lambda^1}{E_\pi}  \, H_{21}^1
- H_{21}^1 
\frac{\lambda^1}{E_\pi}  H_{22}^2 \, \frac{\lambda^1}{E_\pi^2}  \, H_{21}^1
\bigg] \eta\,, \nn
S_4 &=& \eta  \bigg[ H_{22}^2
\frac{\lambda^2}{E_\pi^2}  H_{21}^1 \, \frac{\lambda^1}{E_\pi}  \, H_{21}^1
- H_{21}^1 
\frac{\lambda^1}{E_\pi}  H_{21}^1 \, \frac{\lambda^2}{E_\pi^2}  \, H_{22}^2
\bigg] \eta\,, \nn
S_5 &=& \eta  \bigg[ H_{22}^2 
\frac{\lambda^2}{E_\pi}  H_{21}^1 \, \frac{\lambda^1}{E_\pi^2}  \, H_{21}^1
- H_{21}^1 
\frac{\lambda^1}{E_\pi^2}  H_{21}^1 \, \frac{\lambda^2}{E_\pi}  \, H_{22}^2
\bigg] \eta\,.
\eeqa
The operators $S_3$, $S_4$ and $S_5$ are the only time-reversal invariant
anti-hermitian operators that can be
constructed out of two vertices $H_{21}^1$ and one Weinberg-Tomozawa vertex
$H_{22}^2$ with $E_\pi$'s appearing only in the
denominators. 
The corresponding $\alpha_i$-dependent class-II contributions to the effective Hamilton
operator  read:
\beqa
\label{class2_add}
\delta  V^{(4)} &=&  [ H^{(0)}, \; S ] \nn
&=& - \alpha_3 \, \eta \bigg[  
 H_{21}^1 \frac{\lambda^1}{E_\pi}  H_{21}^1 \, \eta \,  H_{21}^1 
\frac{\lambda^1}{E_\pi^2}  H_{22}^2 \, \frac{\lambda^1}{E_\pi}  \, H_{21}^1 - 
H_{21}^1 \frac{\lambda^1}{E_\pi}  H_{21}^1 \, \eta \,  H_{21}^1 
\frac{\lambda^1}{E_\pi}  H_{22}^2 \, \frac{\lambda^1}{E_\pi^2}  \, H_{21}^1
 \bigg] \eta \nn
&& {} - \alpha_4 \, \eta \bigg[  
 H_{21}^1 \frac{\lambda^1}{E_\pi}  H_{21}^1 \, \eta \,  H_{22}^2 
\frac{\lambda^2}{E_\pi^2}  H_{21}^1 \, \frac{\lambda^1}{E_\pi}  \, H_{21}^1
- H_{21}^1 \frac{\lambda^1}{E_\pi}  H_{21}^1 \, \eta \, H_{21}^1 
\frac{\lambda^1}{E_\pi}  H_{21}^1 \, \frac{\lambda^2}{E_\pi^2}  \, H_{22}^2
 \bigg] \eta \\
&& {} - \alpha_5 \, \eta \bigg[  
 H_{21}^1 \frac{\lambda^1}{E_\pi}  H_{21}^1 \, \eta \, H_{22}^2 
\frac{\lambda^2}{E_\pi}  H_{21}^1 \, \frac{\lambda^1}{E_\pi^2}  \, H_{21}^1
-   H_{21}^1 \frac{\lambda^1}{E_\pi}  H_{21}^1 \, \eta \, H_{21}^1 
\frac{\lambda^1}{E_\pi^2}  H_{21}^1 \, \frac{\lambda^2}{E_\pi}  \, H_{22}^2
\bigg] \eta  + \mbox{h.c.} + \ldots\,, \nonumber 
\eeqa
where the ellipses refer to terms involving an insertion of either the nucleon kinetic energy
 $H_{20}^2$ or the contact interaction $H_{40}^2$ (class-V contributions). 
The constants $\alpha_3$, $\alpha_4$ and $\alpha_5$ are constrained by the 
requirement that the OPEP factorizes out in the 3NF diagrams shown in
 Fig.~\ref{fig5a}.  
\begin{figure}[tb]
\vskip 1 true cm
\includegraphics[width=8.5cm,keepaspectratio,angle=0,clip]{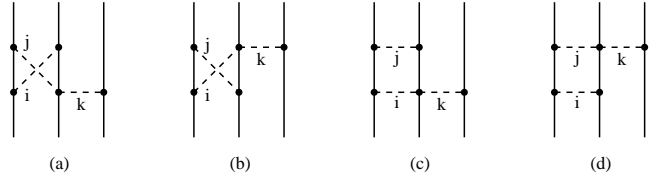}
    \caption{
         Class-II three-nucleon diagrams which provide constraints on
         $\alpha_3$, $\alpha_4$ and $\alpha_5$ as explained in the text. For 
         notation, see Figs.~\ref{fig1} and \ref{fig1a}.
\label{fig5a} 
 }
\end{figure}
This guarantees that all ultraviolet divergences arising in the
corresponding
loop integrals are absorbable into a redefinition of the LEC $D$ entering the 3NF 
(g) in Fig.~\ref{fig4a}. We emphasize that the 3NF diagrams shown in
Fig.~\ref{fig5a} are not the only possible class-II diagrams with a single pion
being exchanged between the first two and the third nucleons. 
All other 3NF contributions which are not shown in Fig.~\ref{fig5a} are found
to be proportional to $\omega_k^{-1}$
irrespectively on the values of $\alpha_i$ and, therefore, do not constrain
these constants. 
Evaluating matrix elements of the operators in Eqs.~(\ref{class2}) and
(\ref{class2_add}) for all possible ``time orderings'' of diagrams in
Fig.~\ref{fig5a} yields:
\beqa
\label{match_class2}
\bar V^{\rm a}_{3N} &=& \frac{4}{\omega_i \omega_j \omega_k} \,, \nn
\bar V^{\rm b}_{3N} &=& - \bar V^{\rm a}_{3N} \, \Big|_{\{a,i,j\} \to \{b,j,i\}} \,, \nn
\bar V^{\rm c}_{3N} &=&  4 \frac{\alpha_3 + \alpha_5}{\omega_j \omega_k^2}
+ 4 \frac{1 - 2 \alpha_4 + 4 \alpha_5}{\omega_i \omega_j ( \omega_i + \omega_k)}
- 2 \frac{3 - 2 \alpha_4 + 4 \alpha_5}{\omega_i \omega_j  \omega_k}
- 4 \frac{\alpha_3 + \alpha_5}{\omega_i^2 \omega_j}
\,, \nn
\bar V^{\rm d}_{3N} &=& - \bar V^{\rm c}_{3N}\, \Big|_{\{c,i,j\} \to \{d,j,i\}}  \,. 
\eeqa
In order to simplify the above expressions, we have taken into account 
the pion kinetic energy arising from the time derivative in the
Weinberg-Tomozawa vertex according to:
\beqa
\langle \pi_a (\vec q_1) \, N | H_{22}^2 | \pi_b (\vec q_2 ) \, N \rangle &=& -
(\omega_{\vec q_1} + \omega_{\vec q_2}
) \, v_{ab}\,, \nn
\langle \pi_a  (\vec q_1) \, \pi_b  (\vec q_2) \, N | H_{22}^2 |  N \rangle &=&
- (\omega_{\vec q_1} - \omega_{\vec q_2 } )
\, v_{ab} \,, \nn
\langle  N | H_{22}^2 |  \pi_a  (\vec q_1) \, \pi_b  (\vec q_2) \, N \rangle &=&
(\omega_{\vec q_1} - \omega_{\vec q_2 } )
\, v_{ab} \,.
\eeqa
Here $a$ and $b$ are the pion isospin quantum numbers and 
\beq
v_{ab} = \frac{i}{8 F_\pi^2} \epsilon_{a b c} \tau_c
\frac{1}{\sqrt{\omega_{\vec q_1} \omega_{\vec q_2 }}}\,.
\eeq 
The OPEP between the pair of the first two and the third nucleon factorizes
out in Eq.~(\ref{match_class2}) if the constants $\alpha_i$ fulfill the
relations
\beq
\label{alphas_class2}
\alpha_3 = - \alpha_5\,, \quad \alpha_4 = \frac{1}{2} + 2 \alpha_5\,.
\eeq
With this choice, the expressions for 
$\bar V_{3N}^{\rm c}$ and $\bar V_{3N}^{\rm d}$ from box diagrams take a particularly simple form
and are identical to the ones arising from the cross-box contributions
$\bar V_{3N}^{\rm a}$  and $\bar V_{3N}^{\rm b}$. We further emphasize that although these
constraints do not completely fix the corresponding unitary
transformations, the results for all class-II 3NFs and 4NFs turn out to be
independent on $\alpha_i$. We found that the disconnected diagrams in
Fig.~\ref{fig5} do not yield 4NFs irrespective of the valies of $\alpha_i$. 
Choosing $\alpha_i$ as specified in Eq.~(\ref{alphas_class2}), the last two
diagrams in the first row in Fig.~\ref{fig5} lead to vanishing 4NFs while the
contribution from the first two diagrams in the second row is given by
\beqa
\label{4nf_class2}
V_{\rm Class-II} &=&  \frac{2 g_A^4}{(2 F_\pi)^6} 
\frac{\vec \sigma_1 \cdot \vec q_1 \;\vec \sigma_4 \cdot \vec q_4}{[\vec q_1^{\; 2}  + M_\pi^2]\,
[\vec q_{12}^{\; 2}  + M_\pi^2]\, [\vec q_4^{\; 2}  + M_\pi^2]}  
\, \Big[ (  \fet \tau_1 \cdot \fet \tau_4 \,  \fet \tau_2 \cdot \fet \tau_3
-  \fet \tau_1 \cdot \fet \tau_3 \,  \fet \tau_2 \cdot \fet \tau_4 ) \, \vec q_{12} \cdot \vec q_4 \nn
&& {} +  \fet \tau_1 \times \fet \tau_2 \cdot \fet \tau_4  \; \vec q_{12} \times \vec q_4 \cdot \vec \sigma_3 \Big] +\mbox{all perm.}\,.
\eeqa
The last two diagrams in the second row of Fig.~\ref{fig5} do not involve
reducible topologies. The corresponding
4NFs can, therefore, be evaluated using the Feynman graph techique with no
need to consider all possible ``time-ordered'' diagrams. Using the Feynman 
rule
\beq
\frac{g_A}{4 F_\pi^3} \left[ \tau^a \delta^{bc} \vec \sigma \cdot \left( 4
    \alpha \vec q_1  + (4 \alpha - 1) (\vec q_2 + \vec q_3 ) \right)  + \mbox{
    2 cycl. perm.} \right]\,,
\eeq
for the $\pi \pi \pi NN$ vertex and 
\beqa
\frac{i}{F_\pi^2} \left[ \delta^{ab}  \delta^{cd} \left( \left[ ( q_1 + q_2 )^2 -
    M_\pi^2 \right]  + 2 \alpha \left[ 4 M_\pi^2 - q_1^2 - q_2^2 - q_3^2 - q_4^2
\right] \right)
+ \mbox{ 2 cycl. perm.} \right] \,,
\eeqa
for the $\pi\pi\pi \pi$-vertex, 
where the superscripts refer to the pion isospin quantum numbers and $q_i$
denote the corresponding pion outgoing momenta, 
we obtain the following result:
\beqa
\label{4nf_class2b}
V_{\rm Class-II} &=& \frac{g_A^4}{(2 F_\pi)^6} 
\frac{\vec \sigma_2 \cdot \vec q_2 \;\vec \sigma_3 \cdot \vec q_3 \;
\vec \sigma_4 \cdot \vec q_4}{[\vec q_2^{\;2}  + M_\pi^2]\,
[\vec q_{3}^{\;2}  + M_\pi^2] \, [\vec q_4^{\;2}  + M_\pi^2]} 
\, \fet \tau_1 \cdot \fet \tau_2 \,  \fet \tau_3 \cdot \fet \tau_4 \; 
\Big[ \vec \sigma_1 \cdot (\vec q_3 + \vec q_4 ) - 4 \alpha 
\vec \sigma_1 \cdot (\vec q_2 + \vec q_3 + \vec q_4 ) \Big]
 \nn   [4pt]
&+&   \frac{g_A^4}{2 (2 F_\pi)^6} 
\, \frac{\vec \sigma_1 \cdot \vec q_1 \;\vec \sigma_2 \cdot \vec q_2 \;\vec
  \sigma_3 \cdot  \vec q_3 \;\vec \sigma_4 \cdot \vec q_4}
{[\vec q_1^{\;2}  + M_\pi^2]\,  [\vec q_{2}^{\;2}  + M_\pi^2] \, [\vec
  q_{3}^{\;2}  + M_\pi^2] \, [\vec q_4^{\;2}  + M_\pi^2]} 
\fet \tau_1 \cdot \fet \tau_2 \,  \fet \tau_3 \cdot \fet \tau_4  
\Big[ \left( \vec q_1 + \vec q_2 \, \right)^2 + M_\pi^2\nn [2pt]
&-&2 \alpha \left(  4 M_\pi^2 + \vec q_1^{\;2} + \vec q_2^{\;2} + \vec
  q_3^{\;2} + \vec q_4^{\;2}  \right)
\Big]
 + \mbox{all perm.}\,. 
\eeqa
It is easy to see that terms proportional to $\alpha$ cancel out each other. 
Thus, while the individual contributions from the last two diagrams in
Fig.~\ref{fig5} do depend on the arbitrary constant $\alpha$, 
the $\alpha$-dependence drops out completely in the total result. 
Notice further that the individual  contributions of these two diagrams are
given in Ref.~\cite{Epelbaum:2006eu} for the choice $\alpha = 0$. 

\item
{\bf Class-III} contributions proportional to $g_A^2$.

This class of contributions arises from diagrams with two insertions of the 
Weinberg-Tomozawa vertex $H_{22}^2$ or one insertion of the $\pi \pi NNNN$ vertex $H_{42}^4$
defined in the last line of Eq.~(\ref{ham}), see Fig.~\ref{fig7}. 
\begin{figure}[tb]
\vskip 1 true cm
\includegraphics[width=7.2cm,keepaspectratio,angle=0,clip]{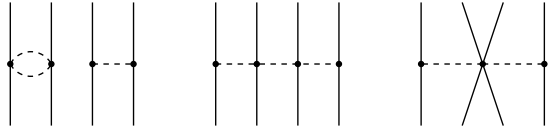}
    \caption{
         Class-III contributions to the 4NF. For 
         notation, see Figs.~\ref{fig1} and \ref{fig1a}.
\label{fig7} 
 }
\end{figure}
The corresponding operators read:  
\beqa
\label{class3}
V^{(4)} &=&  \eta \bigg[ \frac{1}{2}  
  H_{21}^1 \frac{\lambda^1}{E_\pi} H_{21}^1 \, \eta \, H_{22}^2 \frac{\lambda^2}{E_\pi^2} H_{22}^2
\; + \;   \frac{1}{2} H_{21}^1 \frac{\lambda^1}{E_\pi^2} H_{21}^1 \, \eta \, H_{22}^2 \frac{\lambda^2}{E_\pi
  } H_{22}^2 \; 
-\; H_{21}^1 \frac{\lambda^1}{E_\pi} H_{21}^1 \frac{\lambda^2}{E_\pi} H_{22}^2 
\frac{\lambda^2}{E_\pi} H_{22}^2 \nn  [2pt]
&& {} 
- H_{21}^1 \frac{\lambda^1}{E_\pi} H_{22}^2 \frac{\lambda^1}{E_\pi} H_{21}^1 
\frac{\lambda^2}{E_\pi} H_{22}^2\;
- \; \frac{1}{2} H_{21}^1 \frac{\lambda^1}{E_\pi} H_{22}^2 \frac{\lambda^1}{E_\pi} H_{22}^2 
\frac{\lambda^1}{E_\pi} H_{21}^1
\; - \; H_{21}^1 \frac{\lambda^1}{E_\pi} H_{22}^2 \frac{\lambda^3}{E_\pi} H_{21}^1 
\frac{\lambda^2}{E_\pi} H_{22}^2 \nn  [2pt]
&& {} 
- \frac{1}{2}  H_{21}^1 \frac{\lambda^1}{E_\pi} H_{22}^2 \frac{\lambda^3}{E_\pi} H_{22}^2 
\frac{\lambda^1}{E_\pi} H_{21}^1
\; -\;  \frac{1}{2} H_{22}^2 \frac{\lambda^2}{E_\pi} H_{21}^1
\frac{\lambda^3}{E_\pi} H_{21}^1 
\frac{\lambda^2}{E_\pi} H_{22}^2 
\;- \; \frac{1}{2} H_{22}^2 \frac{\lambda^2}{E_\pi} H_{21}^1
\frac{\lambda^1}{E_\pi} H_{21}^1 
\frac{\lambda^2}{E_\pi} H_{22}^2  \nn  [2pt]
&& {} 
 +  H_{21}^1 \frac{\lambda^1}{E_\pi} H_{21}^1
\frac{\lambda^2}{E_\pi} H_{42}^4  \; + \; \frac{1}{2} H_{21}^1 \frac{\lambda^1}{E_\pi} H_{42}^4
\frac{\lambda^1}{E_\pi} H_{21}^1 
\bigg ] \eta + \mbox{h.c.} \,.
\eeqa
The class-III contributions are not affected by the $\eta$-space unitary
  transformations. The disconnected diagram is again found to
  produce vanishing 4NF contribution. The 4NF contributions from the last two diagrams
  derived using Eq.~(\ref{class3}) do not vanish individually but cancel
  each other. This result can also be obtained in a
  simpler way using the Feynman graph technique. In this framework, one only
  has to consider the second graph in Fig.~\ref{fig7}. The last diagram does
  not appear 
  since there is no $\pi \pi NNNN$ vertex of dimension 
$\Delta_i =0$ in the
  Lagrangian. Due to the four-momentum conservation at each vertex and the fact that 
the Weinberg-Tomozawa vertex contains a time derivative of the pion field, the 
contribution of the second Feynman diagram in  Fig.~\ref{fig7} is suppressed
  by $Q^2/m^2$ and, 
therefore, shifted to higher 
orders.

\item
{\bf Class-IV} contributions proportional to $g_A^4 \, C_{S,T}$.

This class of contributions arises from diagrams
involving four vertices $H_{21}^1$ and one insertion of the 
leading-order contact interactions $H_{40}^2$, see Fig.~\ref{fig8}. 
\begin{figure}[tb]
\vskip 1 true cm
\includegraphics[width=16.0cm,keepaspectratio,angle=0,clip]{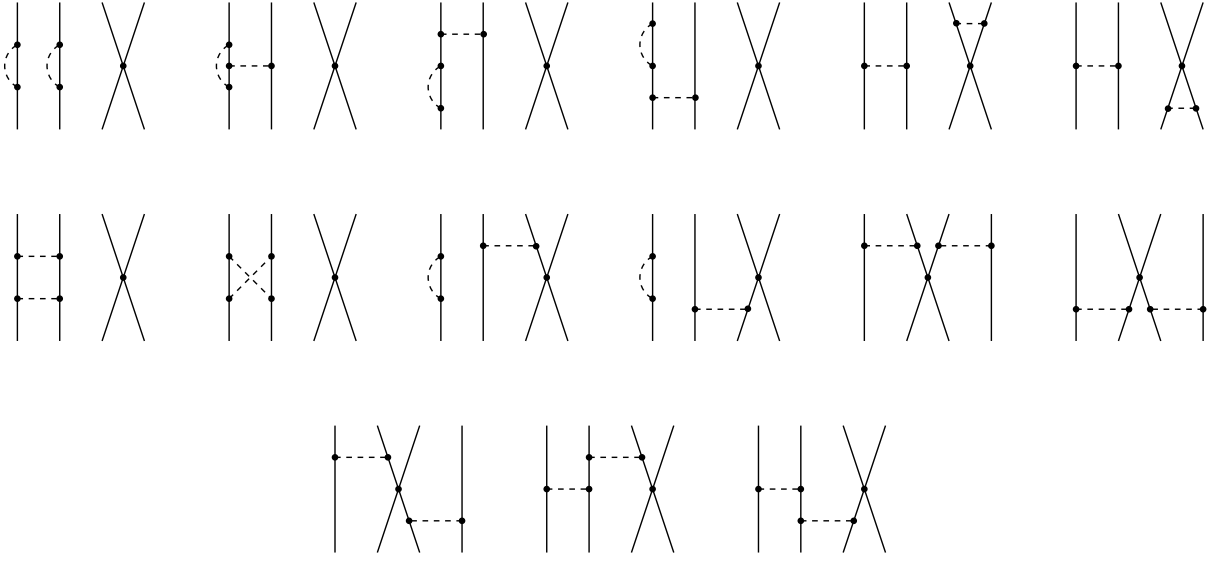}
    \caption{
         Class-IV contributions to the 4NF. For 
         notation, see Figs.~\ref{fig1} and \ref{fig1a}.
\label{fig8} 
 }
\end{figure}
From Eqs.~(\ref{effpot})-(\ref{n13}), one obtains: 
\beqa
\label{class4}
V^{(4)} &=&  \eta \bigg[  - \frac{1}{2}  
  H_{21}^1 \frac{\lambda^1}{E_\pi} H_{21}^1 \, \eta \, H_{21}^1 \frac{\lambda^1}{E_\pi} H_{40}^2
  \frac{\lambda^1}{E_\pi^2} H_{21}^1  
 \; - \; \frac{1}{2}  
  H_{21}^1 \frac{\lambda^1}{E_\pi} H_{21}^1 \, \eta \, H_{21}^1 \frac{\lambda^1}{E_\pi^2} H_{40}^2
  \frac{\lambda^1}{E_\pi} H_{21}^1   
  \nn [2pt]
&& {}
 - \frac{1}{2}  
  H_{21}^1 \frac{\lambda^1}{E_\pi^2} H_{21}^1 \, \eta \, H_{21}^1 \frac{\lambda^1}{E_\pi} H_{40}^2
  \frac{\lambda^1}{E_\pi} H_{21}^1  
\; - \;  \frac{1}{2}  
  H_{40}^2 \, \eta \,  H_{21}^1 \frac{\lambda^1}{E_\pi} H_{21}^1
  \frac{\lambda^2}{E_\pi} H_{21}^1
  \frac{\lambda^1}{E_\pi^2} H_{21}^1   \nn [2pt]
&& {}
- \frac{1}{2}  
  H_{40}^2 \, \eta \,  H_{21}^1 \frac{\lambda^1}{E_\pi} H_{21}^1
  \frac{\lambda^2}{ E_\pi^2} H_{21}^1
  \frac{\lambda^1}{E_\pi} H_{21}^1   
\; - \;  \frac{1}{2}  
  H_{40}^2 \, \eta \,  H_{21}^1 \frac{\lambda^1}{E_\pi^2} H_{21}^1
  \frac{\lambda^2}{E_\pi} H_{21}^1
  \frac{\lambda^1}{E_\pi} H_{21}^1    \nn [2pt]
&& {}
+ \frac{1}{2}  
  H_{40}^2 \, \eta \,  H_{21}^1 \frac{\lambda^1}{E_\pi} H_{21}^1
  \, \eta \,  H_{21}^1
  \frac{\lambda^1}{E_\pi^3} H_{21}^1   
\; + \;  \frac{3}{8}  
  H_{40}^2 \, \eta \,  H_{21}^1 \frac{\lambda^1}{E_\pi^2} H_{21}^1
  \, \eta \,  H_{21}^1
  \frac{\lambda^1}{E_\pi^2} H_{21}^1  \nn [2pt]
&& {}
 + \frac{1}{2}  
  H_{21}^1 \frac{\lambda^1}{E_\pi} H_{21}^1 \, \eta \, H_{40}^2 \, \eta \,  H_{21}^1
  \frac{\lambda^1}{E_\pi^3} H_{21}^1    
\;  + \;  \frac{1}{8}  
  H_{21}^1 \frac{\lambda^1}{E_\pi^2} H_{21}^1 \, \eta \, H_{40}^2 \, \eta \,  H_{21}^1
  \frac{\lambda^1}{E_\pi^2} H_{21}^1    \nn [2pt]
&& {} 
 +   
  H_{21}^1 \frac{\lambda^1}{E_\pi} H_{21}^1 \frac{\lambda^2}{E_\pi}
  H_{21}^1  \frac{\lambda^1}{E_\pi} H_{40}^2 \frac{\lambda^1}{E_\pi} H_{21}^1  
\; + \;  \frac{1}{2} 
  H_{21}^1 \frac{\lambda^1}{E_\pi} H_{21}^1 \frac{\lambda^2}{E_\pi}
  H_{40}^2  \frac{\lambda^2}{E_\pi} H_{21}^1 \frac{\lambda^1}{E_\pi} H_{21}^1  
\bigg ] \eta + \mbox{h.c.} 
\eeqa
In addition to terms listed above one has to include
contributions arising from the $\eta$-space unitary transformation defined in
Eqs.~(\ref{UT}), (\ref{add1}) and acting on $H_{40}^2$,\footnote{These terms are
  not shown explicitly in Eq.~(\ref{class1_add}).}
\beqa
\label{class4_add1}
\delta  V^{(4)} &=& \eta \bigg[ \frac{1}{2} \, H_{21}^1 \frac{\lambda^1}{E_\pi}  H_{21}^1 \,
\eta \, H_{21}^1 \frac{\lambda^1}{E_\pi^3}  H_{21}^1 \, \eta \, H_{40}^2 - 
  \frac{1}{2} \, H_{21}^1 \frac{\lambda^1}{E_\pi^3}  H_{21}^1 \,
\eta \, H_{21}^1 \frac{\lambda^1}{E_\pi}  H_{21}^1 \, \eta \, H_{40}^2 \nn [2pt]
 && - \frac{1}{4} \, H_{21}^1 \frac{\lambda^1}{E_\pi}  H_{21}^1  \frac{\lambda^2}{E_\pi}
 H_{21}^1 \frac{\lambda^1}{E_\pi^2}  H_{21}^1 \, \eta \, H_{40}^2 
+ \frac{1}{4} \, H_{21}^1 \frac{\lambda^1}{E_\pi^2}  H_{21}^1  \frac{\lambda^2}{E_\pi}
 H_{21}^1 \frac{\lambda^1}{E_\pi}  H_{21}^1 \, \eta \, H_{40}^2 \bigg] \eta + \mbox{h.c.} 
\eeqa
Here, we already adopted the values for $\alpha_{1,2}$ from Eq.~(\ref{alpha_1_2}).
Finally, one needs to take into account contributions which arise from yet
unconsidered unitary transformation in Eq.~(\ref{UT}) with
$S = \alpha_6 S_6$ and the generator $S_6$ given by
\beq
\label{S_alpha6}
S_6 = \eta  \bigg[ - H_{21}^1 
\frac{\lambda^1}{E_\pi^3}  H_{21}^1 \, \eta  \, H_{40}^2
+ H_{40}^2\, \eta \,  H_{21}^1 \, \frac{\lambda^1}{E_\pi^3}  \, H_{21}^1
\bigg] \eta\,. 
\eeq
The corresponding contributions to the effective Hamilton operator read:
\beqa
\delta  V^{(4)} &=& \alpha_6 \, \eta \bigg[ H_{21}^1 \frac{\lambda^1}{E_\pi}  H_{21}^1 \,
\eta \, H_{21}^1 \frac{\lambda^1}{E_\pi^3}  H_{21}^1 \, \eta \, H_{40}^2 + 
H_{40}^2  \, \eta \, H_{21}^1 \frac{\lambda^1}{E_\pi^3}  H_{21}^1 \,
\eta \, H_{21}^1 \frac{\lambda^1}{E_\pi}  H_{21}^1  \nn [2pt]
&& {} - H_{21}^1 \frac{\lambda^1}{E_\pi}  H_{21}^1 \,
\eta \, H_{40}^2 \, \eta \, H_{21}^1 \frac{\lambda^1}{E_\pi^3}  H_{21}^1 
- H_{21}^1 \frac{\lambda^1}{E_\pi^3}  H_{21}^1 \,
\eta \, H_{40}^2 \, \eta \, H_{21}^1 \frac{\lambda^1}{E_\pi}  H_{21}^1 \bigg] \eta \,.
\eeqa
Notice that the second possible generator 
\beq
\eta  \bigg[ - H_{21}^1 
\frac{\lambda^1}{E_\pi^2}  H_{40}^2 \frac{\lambda^1}{E_\pi}  H_{21}^1
+ H_{21}^1 \frac{\lambda^1}{E_\pi} H_{40}^2 \frac{\lambda^1}{E_\pi^2}  \, H_{21}^1
\bigg] \eta = 0
\eeq
does not lead to a non-trivial transformation. The constant $\alpha_6$ can be
determined unambigously by the requirement that the OPEP factorizes out in the 
3NF diagrams shown in Fig.~\ref{fig9}.  
\begin{figure}[tb]
\vskip 1 true cm
\includegraphics[width=14.0cm,keepaspectratio,angle=0,clip]{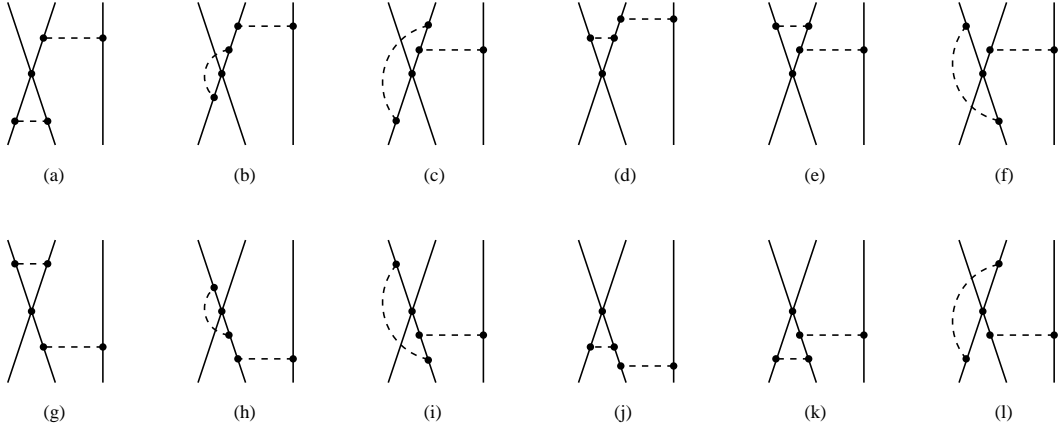}
    \caption{
         Class-IV three-nucleon diagrams which do not involve 
         self-energy insertions. For 
         notation, see Figs.~\ref{fig1} and \ref{fig1a}.
\label{fig9} 
 }
\end{figure}
The energy denominators for the individual graphs in this figure read:
\beqa
\bar V^{\rm a}_{3N} &=& \bar V^{\rm g}_{3N} = 2 (1 - 2 \alpha_6 ) \frac{\omega_i^2 +
  \omega_j^2}{\omega_i^3 \omega_j^3} \,, \nn
\bar V^{\rm b}_{3N} &=& - \bar V^{\rm c}_{3N} =  \bar
V^{\rm h}_{3N} = - \bar
V^{\rm i}_{3N}=  - \frac{2}{\omega_i \omega_j^3} \,, \nn
\bar V^{\rm d}_{3N} &=& \bar V^{\rm j}_{3N} =  \frac{4 \alpha_6}{\omega_i \omega_j^3}
\,, \nn
\bar V^{\rm e}_{3N} &=& \bar V^{\rm k}_{3N} = - \frac{2}{\omega_i \omega_j^3} - 2 (1 - 2
\alpha_6 ) \frac{1}{\omega_i^3 \omega_j} \,, \nn
\bar V^{\rm f}_{3N} &=& \bar V^{\rm l}_{3N} =0\,.
\eeqa
Here, $i$ labels the pion exchanged between the first
two nucleons and the third one while $j$ denotes the pion which does not
interact with the third nucleon. The OPEP factorizes out in the above terms if one
sets 
\beq
\label{alpha6}
\alpha_6 = \frac{1}{2}\,.
\eeq
We have verified that this choice also leads to the desired 
$1/\omega_i$-dependence for the class-IV 3NF contributions involving self-energy
insertions which are not shown in Fig.~\ref{fig9}. 
With the $\eta$-space unitary transformations being fixed as described above
only the last two diagrams in Fig.~\ref{fig8} yield non-vanishing 4NF 
contributions: 
\beq
\label{4nf_class4}
V_{\rm Class-IV} = 4 C_T \frac{g_A^4}{(2 F_\pi)^4} \,
\frac{\vec \sigma_1 \cdot \vec q_1 \; \vec \sigma_3 \times \vec \sigma_4  \cdot \vec q_{12}}
{[\vec q_1^{\;2}  + M_\pi^2]\, [\vec q_{12}^{\;2}  + M_\pi^2]^2} 
\,   \Big[ \fet \tau_1 \cdot \fet \tau_3 \; \vec q_1 \times \vec q_{12} \cdot \vec \sigma_2   - 
\fet \tau_1 \times \fet \tau_2 \cdot \fet \tau_3 \; \vec q_{1} \cdot \vec q_{12} \Big] 
 +  \mbox{all perm.}. \nn [4pt]
\eeq

\item
{\bf Class-V} contributions proportional to $g_A^2 \, C_{S,T}$.

The class-V contributions arise from diagrams constructed from 
two vertices $H_{21}^1$, one Weinberg-Tomozawa vertex $H_{22}^2$ and the 
leading contact interaction $H_{40}^2$, see Fig.~\ref{fig10}. 
\begin{figure}[tb]
\vskip 1 true cm
\includegraphics[width=14.0cm,keepaspectratio,angle=0,clip]{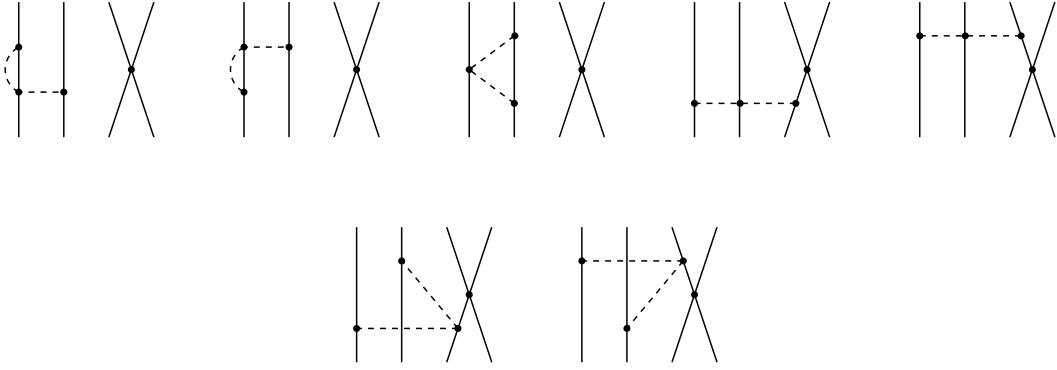}
    \caption{
         Class-V contributions to the 4NF. For 
         notation, see Figs.~\ref{fig1} and \ref{fig1a}.
\label{fig10} 
 }
\end{figure}
From Eqs.~(\ref{effpot})-(\ref{n13}), one obtains: 
\beqa
\label{class5}
V^{(4)} &=&  \eta \bigg[  \frac{1}{2}  
  H_{40}^2 \, \eta \, H_{21}^1 \frac{\lambda^1}{E_\pi} H_{21}^1 \frac{\lambda^2}{E_\pi^2} H_{22}^2
\; + \;  
\frac{1}{2}  
  H_{40}^2 \, \eta \, H_{21}^1 \frac{\lambda^1}{E_\pi} H_{22}^2 \frac{\lambda^1}{E_\pi^2} H_{21}^1
\; + \;  
\frac{1}{2}  
  H_{40}^2 \, \eta \, H_{22}^2 \frac{\lambda^2}{E_\pi} H_{21}^1
  \frac{\lambda^1}{E_\pi^2} H_{21}^1 \nn
&& {} 
 + \frac{1}{2}  H_{40}^2 \, \eta \, H_{21}^1 \frac{\lambda^1}{E_\pi^2} H_{21}^1 \frac{\lambda^2}{E_\pi} H_{22}^2
\; + \;  
\frac{1}{2}  
  H_{40}^2 \, \eta \, H_{21}^1 \frac{\lambda^1}{E_\pi^2} H_{22}^2 \frac{\lambda^1}{E_\pi} H_{21}^1
\; + \;  
\frac{1}{2}  
  H_{40}^2 \, \eta \, H_{22}^2 \frac{\lambda^2}{E_\pi^2} H_{21}^1
  \frac{\lambda^1}{E_\pi} H_{21}^1 \\
&& {} 
 - H_{21}^1 \frac{\lambda^1}{E_\pi}  H_{21}^1 \frac{\lambda^2}{E_\pi} H_{40}^2 \frac{\lambda^2}{E_\pi} H_{22}^2
 - H_{21}^1 \frac{\lambda^1}{E_\pi}  H_{22}^2 \frac{\lambda^1}{E_\pi} H_{40}^2 \frac{\lambda^1}{E_\pi} H_{21}^1
 - H_{21}^1 \frac{\lambda^1}{E_\pi}  H_{40}^2 \frac{\lambda^1}{E_\pi} H_{21}^1 \frac{\lambda^2}{E_\pi} H_{22}^2
\bigg ] \eta + \mbox{h.c.}\,. \nonumber
\eeqa
In addition, one has to take into account the contributions arising from the $\eta$-space unitary
  transformations proportional to $\alpha_{3,4,5}$ and acting on $H_{40}^2$:
\beqa
\label{class5_add}
\delta  V^{(4)} &=&  - \alpha_5 \, \eta \bigg[  
 H_{40}^2  \, \eta \,  H_{21}^1 
\frac{\lambda^1}{E_\pi^2}  H_{22}^2 \, \frac{\lambda^1}{E_\pi}  \, H_{21}^1 - 
H_{40}^2 \, \eta \,  H_{21}^1 
\frac{\lambda^1}{E_\pi}  H_{22}^2 \, \frac{\lambda^1}{E_\pi^2}  \, H_{21}^1
 \bigg] \eta \nn
&& {} + \left( \frac{1}{2} + 2 \alpha_5  \right) \, \eta \bigg[  
 H_{40}^2 \, \eta \,  H_{22}^2 
\frac{\lambda^2}{E_\pi^2}  H_{21}^1 \, \frac{\lambda^1}{E_\pi}  \, H_{21}^1
- H_{40}^2 \, \eta \, H_{21}^1 
\frac{\lambda^1}{E_\pi}  H_{21}^1 \, \frac{\lambda^2}{E_\pi^2}  \, H_{22}^2
 \bigg] \eta \nn
&& {} + \alpha_5 \, \eta \bigg[  
 H_{40}^2  \, \eta \, H_{22}^2 
\frac{\lambda^2}{E_\pi}  H_{21}^1 \, \frac{\lambda^1}{E_\pi^2}  \, H_{21}^1
-   H_{40}^2 \, \eta \, H_{21}^1 
\frac{\lambda^1}{E_\pi^2}  H_{21}^1 \, \frac{\lambda^2}{E_\pi}  \, H_{22}^2
\bigg] \eta  + \mbox{h.c.} \,.  
\eeqa
Here, we have adopted the values for $\alpha_{3,4}$ from
 Eq.~(\ref{alphas_class2}). Similar to the class-II forces, the
 resulting 4NF turns out to be $\alpha_5$-independent. From the diagrams shown
 in Fig.~\ref{fig10}, only the last two in the first row generate the
 non-vanishing contribution which reads:
\beq
\label{4nf_class5}
V_{\rm Class-V} = 2 C_T \frac{g_A^2}{(2 F_\pi)^4} \,
\frac{\vec \sigma_1 \cdot \vec q_1 \; \vec \sigma_3 \times \vec \sigma_4
 \cdot \vec  q_{12}} {[\vec q_1^{\;2}  + M_\pi^2] \, [\vec q_{12}^{\;2}  +
 M_\pi^2]} \; \fet \tau_1 \times \fet \tau_2 \cdot \fet \tau_3  
+ \mbox{all perm.}. 
\eeq

\item
{\bf Class-VI} contributions proportional to $C_{S,T}$.

The only way the class-VI contributions can be generated is from a single disconnected diagram
shown in Fig.~\ref{fig11}. 
\begin{figure}[tb]
\vskip 1 true cm
\includegraphics[width=1.85cm,keepaspectratio,angle=0,clip]{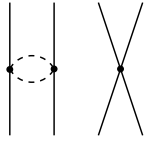}
    \caption{
         Class-VI contribution to the 4NF. For 
         notation, see Figs.~\ref{fig1} and \ref{fig1a}.
\label{fig11} 
 }
\end{figure}
The corresponding terms in the effective potential read: 
\beq
\label{class6}
V^{(4)} = \eta \bigg[ \frac{1}{2} H_{22}^2 \frac{\lambda^2}{E_\pi} H_{40}^2
\frac{\lambda^2}{E_\pi}  H_{22}^2 - 
 \frac{1}{2} H_{22}^2 \frac{\lambda^2}{E_\pi^2} H_{22}^2
\, \eta \,   H_{40}^2 
\bigg ] \eta + \mbox{h.c.}\,. 
\eeq
It is easy to verify that the diagram in Fig.~\ref{fig11} leads to a vanishing  
4NF.  

\item
{\bf Class-VII} contributions proportional to $g_A^2 \, C_{S,T}^2$.

The class-VII contributions arise from diagrams
involving two vertices $H_{21}^1$ and two insertions of the 
leading-order contact interactions $H_{40}^2$, see Fig.~\ref{fig12}. 
From Eqs.~(\ref{effpot})-(\ref{n13}), one obtains: 
\beqa
\label{class7}
V^{(4)} &=& \eta \bigg[  - \frac{1}{2}  
  H_{21}^1 \frac{\lambda^1}{\omega} H_{40}^2  \frac{\lambda^1}{\omega}  H_{40}^2 \frac{\lambda^1}{\omega} H_{21}^1
+   \frac{1}{2} H_{21}^1  \frac{\lambda^1}{\omega} H_{40}^2  \frac{\lambda^1}{\omega^2}  H_{21}^1
\, \eta \,  H_{40}^2
+   \frac{1}{2} H_{21}^1  \frac{\lambda^1}{\omega^2} H_{40}^2  \frac{\lambda^1}{\omega}  H_{21}^1
\, \eta \,  H_{40}^2 \nn
&& {} -   \frac{1}{2} H_{21}^1  \frac{\lambda^1}{\omega^3} H_{21}^1  \, \eta \,  H_{40}^2
\, \eta \,  H_{40}^2
\bigg ] \eta + \mbox{h.c.} \,.
\eeqa
\begin{figure}[tb]
\vskip 1 true cm
\includegraphics[width=16.0cm,keepaspectratio,angle=0,clip]{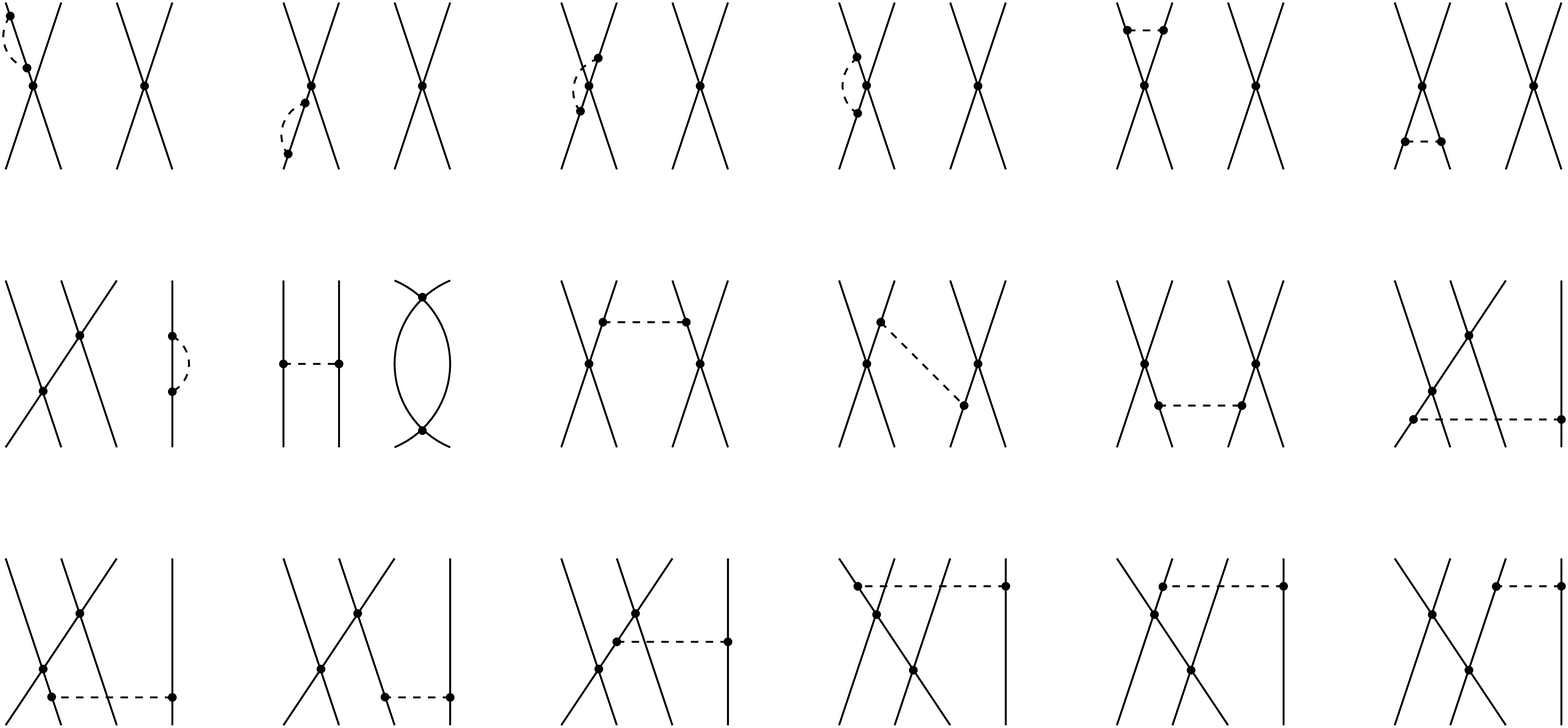}
    \caption{
         Class-VII contributions to the 4NF. For 
         notation, see Figs.~\ref{fig1} and \ref{fig1a}.
\label{fig12} 
 }
\end{figure}
Further terms arise from the $\eta$-space unitary transformation proportional
  to $\alpha_6$, see Eq.~(\ref{S_alpha6}), and acting on $H_{40}^2$. Using the
  value for $\alpha_6$ from Eq.~(\ref{alpha6}), these contributions take the form:
\beq
\delta  V^{(4)} =  \eta \bigg[ - H_{40}^2 \,
\eta \, H_{21}^1 \frac{\lambda^1}{E_\pi^3}  H_{21}^1 \, \eta \, H_{40}^2 
+ \frac{1}{2} H_{40}^2 \,
\eta \, H_{40}^2 \, \eta \, H_{21}^1 \frac{\lambda^1}{E_\pi^3}  H_{21}^1 
+ \frac{1}{2} H_{21}^1 \frac{\lambda^1}{E_\pi^3}  H_{21}^1 \,
\eta \, H_{40}^2 \, \eta \, H_{40}^2 \bigg] \eta \,.
\eeq
From all diagrams shown in Fig.~\ref{fig12}, only the third, fourth and fifth
  ones in the second row lead to the nonvanishing 4NF:
\beq
\label{4nf_class7}
V_{\rm Class-VII} =  2 C_T^2 \frac{g_A^2}{(2 F_\pi)^2} \,
\frac{\vec \sigma_1 \times \vec \sigma_2  \cdot \vec q_{12} \; \vec \sigma_3 \times \vec \sigma_4  \cdot \vec q_{12}}
{[\vec q_{12}^{\;2}  + M_\pi^2]^2} \;  \fet \tau_2 \cdot \fet \tau_3 
 + \mbox{all perm.} \,.
\eeq

\item
{\bf Class-VIII} contributions from disconnected graphs with insertions of a $\Delta_i = 2$-interaction.

Finally, we need to consider contributions involving insertions of the
$\Delta_i = 2$-interactions $H_{20}^2$, $H_{02}^2$, $H_{21}^3$ and $H_{40}^4$
and arising from disconnected diagrams shown in Fig.~\ref{fig13}. 
The corresponding terms in the effective Hamilton operator are listed
in Appendix \ref{app1}. Evaluating matrix elements of these terms
we have verified that all disconnected diagrams in Fig.~\ref{fig13} lead to
vanishing contributions to the 4NF. 
\end{itemize}

\subsection{Discussion}
\label{sec:disc}

In the previous section we have worked out the leading
4NF which is given in  Eqs.~(\ref{4nf_class1}), (\ref{4nf_class2}), 
(\ref{4nf_class2b}), (\ref{4nf_class4}),  (\ref{4nf_class5}) and
(\ref{4nf_class7}).  
Some of these contributions have, in fact, already been considered in the past. In
particular, McManus and Riska \cite{Mcmanus:1980ep} discussed a long time ago
the 4NF generated by the
last two diagrams in Fig.~\ref{fig5}. These terms were also studied by
Robilotta \cite{Robilotta:1985gv} who, in addition, considered effects due to
intermediate delta excitations and exchange of rho and axial-vector
mesons. In that work, the $\pi \pi \pi NN$ and  $\pi \pi \pi \pi$ vertices
were parametrized in terms of the constant $\xi$ which plays a role similar to
$\alpha$ in Eq.~(\ref{alfa_def}). 
Our result for the penultimate diagram in Fig.~\ref{fig5} 
shown in the first line of Eq.~(\ref{4nf_class2b}) agrees with
the one of Ref.~\cite{Mcmanus:1980ep} if one chooses $\alpha = 1/4$ and with the
one of Ref.~\cite{Robilotta:1985gv} if one chooses $4 \alpha = \xi$.  
The result for the last diagram in Fig.~\ref{fig5} which is given in the last two lines 
of Eq.~(\ref{4nf_class2b}) does, however, only agree with the one of
Ref.~\cite{Robilotta:1985gv} if one sets $\alpha =
\xi = 0$, see Eq.~(43) of that work. This disagreement can be traced back to the 
(slightly) different Feynman rules used for the  $\pi \pi \pi \pi$ vertex in
that work.  
Notice further that the $\xi$-dependence does not drop out completely in the
expressions for the 4NF of  Ref.~\cite{Robilotta:1985gv}. All other
nonvanishing 4NF contributions discussed in the previous section result from
diagrams involving reducible topologies and have, to the best of our
knowledge, not been considered before. 

\begin{figure}[tb]
\vskip 1 true cm
\includegraphics[width=16.0cm,keepaspectratio,angle=0,clip]{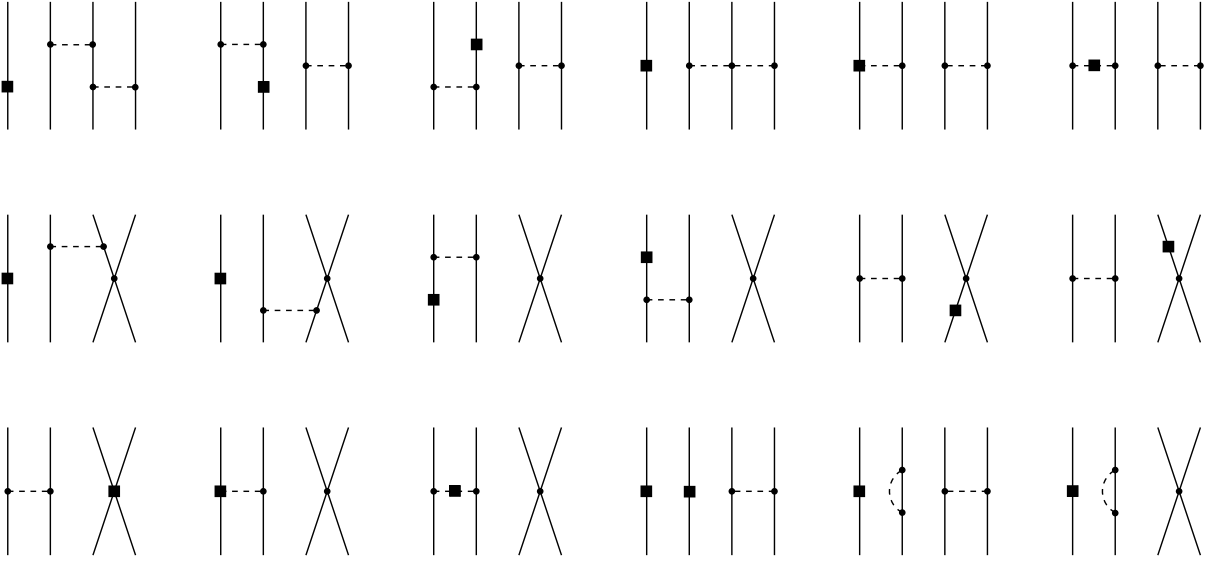}
    \caption{
         Class-VIII contributions to the 4NF. Solid squares denote vertices
         of order $\nu = 4$. For the
         remaining notation see Figs.~\ref{fig1} and \ref{fig1a}.
\label{fig13} 
 }
\end{figure}

In order to test the effects of the 4NFs in few-nucleon systems explicit
calculations will need   
to be performed. To get a rough idea about the size of the 4NF contributions
to e.g.~the $\alpha$-particle  binding energy,
one can look at the strength of the corresponding $r$-space potentials which
are given in appendix \ref{app:coord}. The strengths of the contributions which
do not involve contact interactions is given by
\beq
\frac{g_A^6 M_\pi^7}{(16 \pi F_\pi^2 )^3} \sim 50 \mbox{ keV}
\quad \mbox{and} \quad
\frac{g_A^4 M_\pi^7}{(16 \pi F_\pi^2 )^3} \sim 35 \mbox{ keV}\,,
\eeq
see also \cite{Robilotta:1985gv}. These numbers set the typical scale for the expectation values of
the corresponding operators. Recently, a more careful estimation of the
4NF effects in $^4$He  was carried out by Rozpedzik et
al.~\cite{Rozpedzik:2006yi} which, however, still involves severe
approximations in order to simplify the calculations. 
Expectation values of various individual long-range
4NF contributions (i.e. the ones which do not involve contact interactions) obtained in
that work agree reasonably  well with the
above estimation. For example, for the Gaussian wave function their
magnitudes lie in the range $\sim 70 \ldots 200$ keV. Interestingly, one observes
strong cancellations among various contributions. The cancellation
(to a large extent) between
the contributions corresponding to the last two diagrams in Fig.~\ref{fig5} 
was already mentioned in
Ref.~\cite{Mcmanus:1980ep}. Even more surprising, one observes a nearly complete
cancellation for the total long-range contribution $V_{\rm
  Class-I} + V_{\rm Class-II}$.\footnote{Expressions for $V^c$ and $V^l$ in
  Refs.~\cite{Epelbaum:2006eu} and \cite{Rozpedzik:2006yi} should be taken
  with opposite sign.} 
For all  three wave
functions employed in Ref.~\cite{Rozpedzik:2006yi}, its expectation value 
does not exceed 4 keV.  Further, a similar cancellation occurs for the
class-IV and class-V contributions. For the Gaussian wave function, the
corresponding expectation values are \cite{Rozpedzik:2006yi}
\beq
\langle \Psi_{^4He} | V_{\rm Class-IV} | \Psi_{^4He} \rangle + 
\langle \Psi_{^4He} | V_{\rm Class-V} | \Psi_{^4He} \rangle  = ( 14.2  \mbox{
  keV} ) \times C_T -
(14.9  \mbox{  keV} ) \times C_T \,.
\eeq
Here the value for the LEC $C_T$ should be taken in units of GeV$^{-2}$. 
The LEC $C_T$ has been determined from fits to NN phase
shifts. Its value depends on the cutoffs employed and is typically of the order 
$| C_T | \sim 10$ GeV$^{-2}$ \cite{Epelbaum:2005pn}. Values obtained
from resonance saturation using various phenomenological potentials are of a
similar size \cite{Epelbaum:2001fm}. One therefore concludes that the
contribution to the $^4$He binding energy from 4NF terms proportional to $C_T$
is rather small. The most significant contribution is produced by the class-VII 4NF
which is quadratic in $C_T$.  For the Gaussian wave function it is found to be
\cite{Rozpedzik:2006yi}
\beq
\langle \Psi_{^4He} | V_{\rm Class-VII} | \Psi_{^4He} \rangle  = ( -1.1 \mbox{
  keV} ) \times C_T^2 \,.
\eeq
For $| C_T | \sim 10$ GeV$^{-2}$ the 4NF is, therefore, expected to produce
$\sim 100$ keV attraction in $^4$He. We, however, emphasize that the
expectation values for the contributions involving contact interactions might be
significantly overestimated due to approximations made in
Ref.~\cite{Rozpedzik:2006yi} for the  $^4$He wave functions \cite{Nogga:2007priv}.

\section{Summary and conclusions}
\def\theequation{\arabic{section}.\arabic{equation}}
\label{sec:summary}

In this paper we have analyzed the next-to-next-to-next-to-leading
contribution to the nuclear Hamiltonian in chiral effective
field theory using the method of unitary transformation. 
The pertinent results can be summarized as follows:
\begin{itemize}
\item[i)]
We have presented a new formulation of the chiral power counting which is
particularly suitable for applications based on algebraic approaches such as
the method of unitary transformation. It allowed us to express the recursive
(formal) solution of the decoupling equation in a compact form and to drastically 
simplify the derivation of the effective Hamiltonian as compared to the
formalism given in Ref.~\cite{Epelbaum:1998ka}. 
\item[ii)]
Using this new formulation we have worked out the N$^3$LO contribution
to the effective nuclear Hamiltonian in chiral EFT which involves 2N, 3N and 4N operators. 
The 4N operators yield the leading contribution to the 4NF. 
\item[iii)]
We employed a large class of additional $\eta$-space unitary transformations  
that affect N$^3$LO terms in the effective Hamiltonian. The appearance of such
transformations is a new feature at this order in the chiral
expansion. We found that certain classes of the 3NF at N$^3$LO cannot be
renormalized for an arbitrary choice of the $\eta$-space
transformations. The renormalizability requirement leads to  
constraints for these additional unitary transformations but does not fix them
completely.  
\item[iv)]
We have worked out the leading 4NF both in momentum and configuration
spaces. Enforcing the above mentioned renormalizability constraints we
found no umbiguity in the 4NF due to the employed $\eta$-space unitary
transformations.  
The resulting 4NF is local and depends on the pion decay constant
$F_\pi$, the axial pion-nucleon coupling $g_A$ and the lowest-order NN
contact interaction whose strength $C_T$
is determined in the two--nucleon system \cite{Entem:2003ft,Epelbaum:2004fk}. 
\end{itemize}

In the future, the effects of the 4NF should be tested in the four-nucleon
continuum and in the spectra of light nuclei. Further, it is also important to study
effects due to intermediate delta excitations which are expected to be
significant \cite{Robilotta:1985gv}.   Work along these lines is in progress.

\section*{Acknowledgements}

We are grateful to Hermann Krebs and Ulf-G.~Mei{\ss}ner for helpful
discussions and useful comments on the manuscript.  
This work was supported by the 
Helmholtz Association under the contract number VH-NG-222.

\appendix

\def\theequation{\Alph{section}.\arabic{equation}}
\setcounter{equation}{0}
\section{Contributions of the order $\Delta_i =2$ vertices to the effective
  Hamilton operator at order $\nu = 4$}
\label{app1}

In this appendix we consider the operator structure of 
the effective Hamiltonian at order $\nu = 4$ which results from higher-order
vertices. As already pointed out, these terms lead to disconnected
diagrams which are shown in Fig.~\ref{fig13}. As a general rule, such
disconnected diagrams do not yield nonvanishing contributions to the effective
Hamilton operator in the method of unitary transformation. For the sake of
completeness, we list below the complete operator structure of the class-VIII
contributions. The
relevant vertices at order $\Delta_i = 2$ are $H_{20}^2$, $H_{02}^2$, $H_{21}^3$ and
$H_{40}^4$. The interaction $H_{02}^2$ together with the corresponding pion
tadpole diagrams renormalize the pion mass and wave function. 
As discussed in \cite{Epelbaum:2002gb}, the $H_{02}^2$ vertex does not show up provided the
Lagrangian is formulated in terms of renormalized pion fields and normal
ordering is applied. We will, therefore, not consider the contributions
involving $H_{02}^2$ in what follows. 

Let us begin with the contributions which involve insertions of the nucleon
kinetic energy 
$H_{20}^2$. Using the values for $\alpha_i$ from Eqs.~(\ref{alpha_1_2}), 
(\ref{alphas_class2}) and (\ref{alpha6}) one obtains
\begin{itemize}
\item
terms $\propto g_A^4$:
\beqa
V &=& \eta \bigg[
\frac{1}{2} \,  H_{21}^1 \frac{\lambda^1}{E_\pi}  H_{21}^1 \, \eta \, 
      H_{21}^1 \frac{\lambda^1}{E_\pi^3}  H_{21}^1 \, \eta \, H_{20}^2 
\; -\;   H_{21}^1 \frac{\lambda^1}{E_\pi}  H_{21}^1 \, \eta \, 
      H_{21}^1 \frac{\lambda^1}{E_\pi^3}  H_{20}^2 \, \lambda^1 \,  H_{21}^1 \nn
&& {} + \; \frac{1}{2} \,  H_{21}^1 \frac{\lambda^1}{E_\pi}  H_{21}^1 \, \eta \, 
      H_{20}^2 \, \eta \, H_{21}^1 \frac{\lambda^1}{E_\pi^3}  H_{21}^1 
\; - \; \frac{3}{4} \,  H_{21}^1 \frac{\lambda^1}{E_\pi}  H_{21}^1 \frac{\lambda^2}{E_\pi}  
      H_{21}^1 \frac{\lambda^1}{E_\pi^2}  H_{21}^1 \, \eta \,  H_{20}^2 \nn
&& {} + \; H_{21}^1 \frac{\lambda^1}{E_\pi}  H_{21}^1 \frac{\lambda^2}{E_\pi}  
      H_{21}^1 \frac{\lambda^1}{E_\pi^2}  H_{20}^2 \,  \lambda^1 \, H_{21}^1 
\;  - \; \frac{1}{2} \,  H_{21}^1 \frac{\lambda^1}{E_\pi}  H_{21}^1 \frac{\lambda^2}{E_\pi^2}  
      H_{21}^1 \frac{\lambda^1}{E_\pi}  H_{21}^1 \, \eta \,  H_{20}^2 \nn
&& {} + \; \frac{1}{2} \, H_{21}^1 \frac{\lambda^1}{E_\pi}  H_{21}^1 \frac{\lambda^2}{E_\pi^2}  
      H_{20}^2 \, \lambda^2 \, H_{21}^1 \frac{\lambda^1}{E_\pi}  H_{21}^1 
\; + \; \frac{3}{8} \,  H_{21}^1 \frac{\lambda^1}{E_\pi^2}  H_{21}^1 \, \eta \,   
      H_{21}^1 \frac{\lambda^1}{E_\pi^2}  H_{21}^1 \, \eta \,  H_{20}^2 \nn
&& {} - \frac{1}{2} \, H_{21}^1 \frac{\lambda^1}{E_\pi^2}  H_{21}^1 \, \eta \, 
      H_{21}^1 \frac{\lambda^1}{E_\pi^2}  H_{20}^2 \, \lambda^1 \,  H_{21}^1 
\; + \; \frac{1}{8} \,  H_{21}^1 \frac{\lambda^1}{E_\pi^2}  H_{21}^1 \, \eta \, 
      H_{20}^2 \, \eta \, H_{21}^1 \frac{\lambda^1}{E_\pi^2}  H_{21}^1 \nn
&& {} - \frac{1}{4} \, H_{21}^1 \frac{\lambda^1}{E_\pi^2}  H_{21}^1 \frac{\lambda^2}{E_\pi} 
      H_{21}^1 \frac{\lambda^1}{E_\pi^2}  H_{21}^1 \,\eta \,   H_{20}^2
\bigg] \eta +  \mbox{h.c.} \,,
\eeqa
\item
terms $\propto g_A^2$:
\beqa
V &=& \eta \bigg[ (1 + 2 \alpha_5 ) \, 
      H_{21}^1 \frac{\lambda^1}{E_\pi}  H_{21}^1 \frac{\lambda^2}{E_\pi^2}
      H_{22}^2 \, \eta \, H_{20}^2
\; - \;  H_{21}^1 \frac{\lambda^1}{E_\pi}  H_{21}^1 \frac{\lambda^2}{E_\pi^2} 
      H_{20}^2 \, \lambda^2 \, H_{22}^2  \nn
&& {}
+  \frac{1}{2} (1 - 2 \alpha_5 ) \, H_{21}^1 \frac{\lambda^1}{E_\pi}
      H_{22}^2 \frac{\lambda^1}{E_\pi^2}  H_{21}^1 \, \eta \, H_{20}^2
\; - \; H_{21}^1 \frac{\lambda^1}{E_\pi} H_{22}^2 \frac{\lambda^1}{E_\pi^2}
H_{20}^2 \, \lambda^1 \, H_{21}^1  \nn
&& {}
+  \frac{1}{2} (1 + 2 \alpha_5 ) \, H_{21}^1 \frac{\lambda^1}{E_\pi^2}  H_{21}^1
      \frac{\lambda^2}{E_\pi}  H_{22}^2 \, \eta \, H_{20}^2
\; + \;  \frac{1}{2} (1 + 2 \alpha_5 ) \, H_{21}^1 \frac{\lambda^1}{E_\pi^2}
      H_{22}^2 \frac{\lambda^1}{E_\pi}  H_{21}^1 \, \eta \,  H_{20}^2  \nn
&& {} - H_{21}^1 \frac{\lambda^1}{E_\pi^2}
      H_{20}^2 \, \lambda^1 \,   H_{21}^1 \frac{\lambda^2}{E_\pi}  H_{22}^2 
\; - \;  2 \alpha_5 \,  H_{22}^2 \frac{\lambda^2}{E_\pi^2}  H_{21}^1 
 \frac{\lambda^1}{E_\pi}  H_{21}^1 \, \eta \, H_{20}^2  \nn
&& {} +  \frac{1}{2} (1 - 2 \alpha_5 ) \, H_{22}^2 \frac{\lambda^2}{E_\pi}
      H_{21}^1 \frac{\lambda^1}{E_\pi^2}  H_{21}^1 \, \eta \,  H_{20}^2
\; - \;  \frac{1}{2} H_{21}^1 \frac{\lambda^1}{E_\pi^3}
      H_{21}^1 \, \eta \,   H_{20}^2\, \eta \,  H_{20}^2 \nn
&& {} +  H_{21}^1 \frac{\lambda^1}{E_\pi^3} H_{20}^2 \, \lambda^1  H_{21}^1\,
\eta \,  H_{20}^2
\; - \;  \frac{1}{2} H_{21}^1 \frac{\lambda^1}{E_\pi^3}
      H_{20}^2 \, \lambda^1  \, H_{20}^2\, \lambda^1 \,  H_{21}^1 
 \bigg]  \eta +  \mbox{h.c.} \,,
\eeqa
\item
terms $\propto g_A^2 \, C_{S,T}$:
\beqa
V &=& \eta \bigg[ \frac{1}{2}  \, H_{21}^1 \frac{\lambda^1}{E_\pi} 
H_{40}^2 \frac{\lambda^1}{E_\pi^2}  H_{21}^1  \,  \eta \, H_{20}^2  
\; - \;   H_{21}^1 \frac{\lambda^1}{E_\pi} H_{40}^2 \frac{\lambda^1}{E_\pi^2}
H_{20}^2  \, \lambda^1 \,  H_{21}^1  
\;   + \; \frac{1}{2}  \, H_{21}^1 \frac{\lambda^1}{E_\pi^2} 
H_{40}^2 \frac{\lambda^1}{E_\pi}  H_{21}^1  \, \eta \,  H_{20}^2   \\
&& {} - \frac{1}{2}  \,   H_{21}^1 \frac{\lambda^1}{E_\pi^3} H_{21}^1  \, \eta \,
H_{20}^2  \, \eta \,  H_{40}^2 
\; + \;   H_{21}^1 \frac{\lambda^1}{E_\pi^3} H_{20}^2  \, \lambda^1 \, H_{21}^1 \eta \,
H_{40}^2 
\; - \;  \frac{1}{2}  \, H_{40}^2 \, \eta \,  H_{21}^1
\frac{\lambda^1}{E_\pi^3} H_{21}^1  \, \eta \,  
H_{20}^2   \bigg]  \eta +  \mbox{h.c.} \,. \nonumber
\eeqa
\end{itemize}
It should be emphasized that one can construct 
$\eta$-space unitary transformations with the generators depending on
$H_{20}^2$. Such transformations would produce further contributions which are
not considered here.  

Next consider the operators which involve one insertion of the subleading
pion-nucleon vertex $H_{21}^3$. One obtains 
\begin{itemize}
\item
terms $\propto g_A^3$:
\beqa
V &=& \eta \bigg[  \frac{1}{2} \, H_{21}^1 \frac{\lambda^1}{E_\pi} H_{21}^1 \, \eta
\, H_{21}^1 \frac{\lambda^1}{E_\pi^2} H_{21}^3 
\; + \;  \frac{1}{2} \, H_{21}^1 \frac{\lambda^1}{E_\pi} H_{21}^1 \, \eta
\, H_{21}^3 \frac{\lambda^1}{E_\pi^2} H_{21}^1 
\; - \;  H_{21}^1 \frac{\lambda^1}{E_\pi} H_{21}^1  \frac{\lambda^2}{E_\pi} 
 H_{21}^1 \frac{\lambda^1}{E_\pi} H_{21}^3  \\
&& {}  - H_{21}^1 \frac{\lambda^1}{E_\pi} H_{21}^1  \frac{\lambda^2}{E_\pi} 
 H_{21}^3 \frac{\lambda^1}{E_\pi} H_{21}^1 
\; + \;  \frac{1}{2} \, H_{21}^1 \frac{\lambda^1}{E_\pi^2} H_{21}^1 \, \eta
\, H_{21}^1 \frac{\lambda^1}{E_\pi} H_{21}^3 
\; + \;  \frac{1}{2} \, H_{21}^1 \frac{\lambda^1}{E_\pi^2} H_{21}^1 \, \eta
\, H_{21}^3 \frac{\lambda^1}{E_\pi} H_{21}^1 \bigg]  \eta +  \mbox{h.c.} \,,
\nonumber
\eeqa
\item
terms $\propto g_A \, C_{S,T}$:
\beq
V = \eta \bigg[  H_{21}^1 \frac{\lambda^1}{E_\pi} H_{40}^2
\frac{\lambda^1}{E_\pi} H_{21}^3 
\; - \;  \frac{1}{2} \, H_{21}^1 \frac{\lambda^1}{E_\pi^2} H_{21}^3 \, \eta \,
H_{40}^2 
\; - \;  \frac{1}{2} \, H_{21}^3 \frac{\lambda^1}{E_\pi^2} H_{21}^1 \, \eta \,
H_{40}^2 \bigg]  \eta +  \mbox{h.c.} \,.
\eeq
\end{itemize}
Finally, the contributions involving $H_{40}^4$ read:
\beq
V = \eta \bigg[  H_{21}^1 \frac{\lambda^1}{E_\pi} H_{40}^4
\frac{\lambda^1}{E_\pi} H_{21}^1  
\; - \;  \frac{1}{2} \, H_{21}^1 \frac{\lambda^1}{E_\pi^2} H_{21}^1 \, \eta \,
H_{40}^4 
\; - \;  \frac{1}{2} \, H_{40}^4 \, \eta \, H_{21}^1 \frac{\lambda^1}{E_\pi^2}
H_{21}^1 \bigg]  \eta \,.
\eeq
We have verified via explicit calculations that matrix elements of all  
terms listed above and corresponding to 4N diagrams shown in Fig.~\ref{fig13} vanish.


\def\theequation{\Alph{section}.\arabic{equation}}
\setcounter{equation}{0}
\section{Configuration-space representation of the four-nucleon force}
\label{app:coord}

The derived expressions for the 4NF depend only on the momentum transfer of the nucleons and,
therefore, take the local form in the configuration space:
\beq
\langle \vec r_1 {} ' \, \vec r_2 {} ' \, \vec r_3 {} ' \, \vec r_4 {} ' \,
| \,  V_{4N}  \, |  \, \vec r_1  \, \vec r_2  \, \vec r_3 \, \vec r_4 \, \rangle 
= \delta (\vec r_1{}' - \vec r_1 \, ) \,  \delta (\vec r_2{}' - \vec r_2 \, )
\,  \delta (\vec r_3{}' - \vec r_3 \, ) \,  \delta (\vec r_4{}' - \vec r_4 \,
) \, V_{4N}  \,.
\eeq
Clearly, the locality of the 4NF only holds if one uses a regulator function
which depends on momentum transfers $\vec q_i$ as well. In this appendix we
give the configuration-space representation of the 4NF for a general form of a
local regulator. 
 
Depending on the topology, the 4NF is expressed in Eqs.~(\ref{4nf_class1}),
(\ref{4nf_class2}), 
(\ref{4nf_class2b}), (\ref{4nf_class4}), (\ref{4nf_class5}) and
(\ref{4nf_class7}) in
terms of different sets of momenta, namely  $\{ \vec q_2, \,   \vec q_3, \,
\vec q_4 \}$, $\{ \vec q_1, \,   \vec q_{12}, \, \vec q_4 \}$ and 
 $\{ \vec q_1, \,  \vec q_2, \,   \vec q_3, \, \vec q_4 \}$. 
The corresponding configuration-space expressions have the form:
\beqa
\label{coord}
V_{4N} &=&  
\int \, \frac{d^3 q_2}{(2 \pi)^3} \, \frac{d^3 q_3}{(2 \pi)^3} \, \frac{d^3
  q_4}{(2 \pi)^3} \, e^{i \vec q_2 \cdot \vec r_{21}} \, 
e^{i \vec q_3 \cdot \vec r_{31}} \,e^{i \vec q_4 \cdot \vec r_{41}} \,
V_{4N} (\vec q_{2}, \, \vec q_{3}, \, \vec q_{4} )\,, \nn
V_{4N} &=&  
\int \, \frac{d^3 q_1}{(2 \pi)^3} \, \frac{d^3 q_{12}}{(2 \pi)^3} \, \frac{d^4
  q_4}{(2 \pi)^3} \, e^{i \vec q_1 \cdot \vec r_{12}} \, 
e^{i \vec q_{12} \cdot \vec r_{23}} \,e^{i \vec q_4 \cdot \vec r_{43}} \,
V_{4N} (\vec q_{1}, \, \vec q_{12}, \, \vec q_{4} )\,, \nn
V_{4N} &=&  
\int \, d^3 r_0 \, \int \, \frac{d^3 q_1}{(2 \pi)^3} \,  \frac{d^3 q_2}{(2
  \pi)^3} \,  \frac{d^3 q_3}{(2 \pi)^3} \, \frac{d^3
  q_4}{(2 \pi)^3} \,  e^{i \vec q_1 \cdot \vec r_{10}} \, e^{i \vec q_2 \cdot \vec r_{20}} \, 
e^{i \vec q_3 \cdot \vec r_{30}} \,e^{i \vec q_4 \cdot \vec r_{40}} \,
V_{4N} (\vec q_{2}, \, \vec q_{2}, \, \vec q_{3}, \, \vec q_{4} )\,,
\eeqa
where $\vec r_{ij} \equiv \vec r_i - \vec r_j$. Here we have taken
into account the overall factor $(2 \pi )^3 \delta (\vec P{} ' - \vec P
)$ with $\vec P$ and $\vec P '$ being the total initial and final momenta of
the nucleons, which is not shown explicitly in expressions of
sec.~\ref{sec:ambig}. Using Eq.~(\ref{coord}) we obtain:
\beqa
V_{\rm Class-I} &=&  \frac{g_A^6 M_\pi^7}{(16 \pi F_\pi^2 )^3} \, 
\vec \sigma_1 \cdot \vec \nabla_{12} \, \vec \sigma_4 \cdot \vec \nabla_{43} 
\Big[ ( \fet \tau_1 \cdot \fet \tau_4 \,  \fet \tau_2 \cdot \fet \tau_3
-  \fet \tau_1 \cdot \fet \tau_3 \,  \fet \tau_2 \cdot \fet \tau_4 )
\,\vec \nabla_{12} \cdot \vec \nabla_{23} \, \vec \nabla_{43} \cdot \vec  \nabla_{23} \nn
&& {}\mbox{\hskip 1.3 true cm}  
+ \fet \tau_1 \times \fet \tau_2 \cdot \fet \tau_4  \; \vec \nabla_{12} \cdot \vec \nabla_{23}  \; 
\vec \nabla_{23} \times \vec \nabla_{43} \cdot \vec \sigma_3 
 + \fet \tau_1 \times \fet \tau_3 \cdot \fet \tau_4  \; \vec \nabla_{43} \cdot \vec \nabla_{23}  \; 
\vec \nabla_{12} \times \vec \nabla_{23} \cdot \vec \sigma_2 \nn
&& {}\mbox{\hskip 1.3 true cm}
 + \fet \tau_1 \cdot \fet \tau_4 \; \vec \nabla_{23} \times \vec \nabla_{12} \cdot \vec \sigma_2 \;
\vec \nabla_{23} \times \vec \nabla_{43} \cdot \vec \sigma_3  \Big]
U_1 (x_{12}) \, U_2 (x_{23})\, U_1 (x_{43} )
 +\mbox{all perm.}, \nn [2pt]
V_{\rm Class-II} &=&  \frac{2 g_A^4 M_\pi^7}{(16 \pi F_\pi^2 )^3} \, 
\vec \sigma_1 \cdot \vec \nabla_{12} \, \vec \sigma_4 \cdot \vec \nabla_{43} 
\Big[ (  \fet \tau_1 \cdot \fet \tau_4 \,  \fet \tau_2 \cdot \fet \tau_3
-  \fet \tau_1 \cdot \fet \tau_3 \,  \fet \tau_2 \cdot \fet \tau_4 ) \, \vec \nabla_{23} \cdot \vec \nabla_{43} 
 +  \fet \tau_1 \times \fet \tau_2 \cdot \fet \tau_4  \; \vec \nabla_{23}
\times \vec \nabla_{43} \cdot \vec \sigma_3 \Big] \nn 
&& {} \mbox{\hskip 1.3 true cm} \times U_1 (x_{12}) \, U_1
(x_{23})\, U_1 (x_{43} ) \nn
&+&  \frac{g_A^4 M_\pi^7}{(16 \pi F_\pi^2 )^3} \, 
\fet \tau_1 \cdot \fet \tau_2 \,  \fet \tau_3 \cdot \fet \tau_4 \;
\vec \sigma_2 \cdot \vec \nabla_{21} \;\vec \sigma_3 \cdot \vec \nabla_{31}
\;\vec \sigma_4 \cdot \vec  \nabla_{41}
\; \vec \sigma_1 \cdot (\vec \nabla_{31} + \vec \nabla_{41} ) 
\, U_1 (x_{21}) \, U_1 (x_{31})\, U_1 (x_{41} ) \nn
&+&  \frac{g_A^4 M_\pi^7}{8 \pi (16 \pi F_\pi^2 )^3} \, 
\fet \tau_1 \cdot \fet \tau_2 \,  \fet \tau_3 \cdot \fet \tau_4 \, 
( 1 - \vec \nabla_{1}^2 - \vec \nabla_{2}^2 - 2 \vec \nabla_{1} \cdot \vec
\nabla_{2} ) \;
\vec \sigma_1 \cdot \vec \nabla_{1} \;\vec \sigma_2 \cdot \vec \nabla_{2} \;\vec
  \sigma_3 \cdot  \vec \nabla_{3} \;\vec \sigma_4 \cdot \vec \nabla_{4} 
 \nn
&& {} \mbox{\hskip 1.3 true cm} \times \int \, d^3 x_0 \, U_1 (x_{10}) \, U_1
(x_{20})\, U_1 (x_{30} ) \, U_1 (x_{40} ) +\mbox{all perm.}\,, \nn [2pt]
V_{\rm Class-IV} &=&  \frac{2 g_A^4 C_T M_\pi^7}{(16 \pi F_\pi^2 )^2} \, 
\vec \sigma_1 \cdot \vec \nabla_{12} \; \vec \sigma_3 \times \vec \sigma_4  \cdot \vec \nabla_{23}
\,   \Big[ \fet \tau_1 \cdot \fet \tau_3 \; \vec \nabla_{12} \times \vec \nabla_{23} \cdot \vec \sigma_2   - 
\fet \tau_1 \times \fet \tau_2 \cdot \fet \tau_3 \; \vec \nabla_{12} \cdot
\vec \nabla_{23} \Big]   \nn
&& {} \mbox{\hskip 1.3 true cm} \times 
U_1 (x_{12}) \, U_2 (x_{23})\, W (x_{43} ) +\mbox{all perm.}\,, \nn [2pt]
V_{\rm Class-V} &=& - \frac{2 g_A^2 C_T M_\pi^7}{(16 \pi F_\pi^2 )^2} \, 
\fet \tau_1 \times \fet \tau_2 \cdot \fet \tau_3  \; 
\vec \sigma_1 \cdot \vec \nabla_{12} \; \vec \sigma_3 \times \vec \sigma_4
 \cdot \vec  \nabla_{23} \; U_1 (x_{12}) \, U_1 (x_{23})\, W (x_{43} )
 +\mbox{all perm.}\,, \nn [2pt]
V_{\rm Class-VII} &=& - \frac{g_A^2 C_T^2 M_\pi^7}{16 \pi F_\pi^2 } \, 
\fet \tau_2 \cdot \fet \tau_3 \;
\vec \sigma_1 \times \vec \sigma_2  \cdot \vec \nabla_{23} \; \vec \sigma_3 \times \vec \sigma_4  \cdot \vec \nabla_{23}
\; W (x_{12}) \, U_1 (x_{23})\, W (x_{43} ) +\mbox{all perm.}\,. 
\eeqa
Here, $\vec x_i \equiv M_\pi \, \vec r_i$, $\vec x_{ij} \equiv \vec x_i - \vec
x_j$,  $\vec \nabla_{i}$ ($\vec \nabla_{ij}$) acts on $\vec x_{i}$ ($\vec x_{ij}$) and 
$x_{ij} \equiv | \vec x_{ij}  |$. Further, the functions $U_{1,2}$ and $W$ are
defined as:
\beqa
U_1 (x_{ij}) &=& \frac{4 \pi }{M_\pi} \int \frac{d^3 q}{(2 \pi )^3} \, \frac{e^{i
    \vec q \cdot \vec x_{ij}/M_\pi}}{\vec q \, ^2 + M_\pi^2} \, F_1^\Lambda (q)
\stackrel{\Lambda \to \infty}{\longrightarrow} \frac{e^{-x_{ij}}}{x_{ij}}\,, \nn
U_2 (x_{ij}) &=& 8 \pi M_\pi \, \int \frac{d^3 q}{(2 \pi )^3} \, \frac{e^{i
    \vec q \cdot \vec x_{ij}/M_\pi}}{(\vec q \, ^2 + M_\pi^2)^2} \, F_2^\Lambda (q)
\stackrel{\Lambda \to \infty}{\longrightarrow} e^{-x_{ij}}\,, \nn
W (x_{ij}) &=& \frac{1}{M_\pi^3} \,
  \int \frac{d^3 q}{(2 \pi )^3} \, e^{i
    \vec q \cdot \vec x_{ij}/M_\pi} \, F_3^\Lambda (q)
\stackrel{\Lambda \to \infty}{\longrightarrow} \delta^3 (x_{ij}) \,,
\eeqa
where $F_{1,2,3}^\Lambda$ denotes the corresponding  regulator functions.

\setlength{\bibsep}{0.2em}
\bibliographystyle{h-physrev3}
\bibliography{/home/epelbaum/refs_h-elsevier3}

\end{document}